\documentclass[10pt]{iopart}
\usepackage{iopams}
\usepackage{epsf}
\usepackage{euscript}

\renewcommand{\epsilon}{\varepsilon}
\newcommand{\q}{\quad}
\newcommand{\qq}{\qquad}
\newcommand{\nn}{\nonumber}
\newcommand{\nnn}{\nonumber\\}
\newcommand{\ts}{\textstyle}
\newcommand{\mb}[1]{\mathbf{#1}}
\newcommand{\fpint}{=\hspace{-1.2em}\int} 
\newcommand{\tfpint}{{\textstyle=\hspace{-1em}\int}} 
\newcommand{\Si}[1]{{\rm Si}#1}
\newcommand{\bra}[1]{\langle#1|}
\newcommand{\ket}[1]{|#1\rangle}
\newcommand{\Len}{\EuScript{L}}     
\newcommand{\Area}{\EuScript{A}}    
\newcommand{\srange}{\sigma}
\newcommand{\Domain}{\mathcal{D}}
\newcommand{\openDomain}{\smash{\stackrel{_\circ}{\mathcal{D}}}}
\newcommand{\defas}{:=}
\newcommand{\tfrac}[2]{{\textstyle\frac{#1}{#2}}}
\newcommand{\Chi}{{\raisebox{.1ex}{\large$\chi$}}}
\newcommand{\grad}{{\scriptstyle\boldsymbol{\nabla}}}
\newcommand{\epscomb}[1]{\widehat{#1}^{^{\scriptstyle\epsilon}}}
\newcommand{\rvec}{\mathit{\mathbf{r}}}
\newcommand{\tvec}{\mathit{\mathbf{t}}}
\newcommand{\nvec}{\mathit{\mathbf{n}}}
\newcommand{\Gb}{\overline{\rm G}}
\newcommand{\Gt}{\tilde{\rm G}}
\newcommand{\Gn}{{\rm G}^0}
\newcommand{\Gtn}{\tilde{\rm G}^0}
\newcommand{\At}{\tilde{\rm A}}
\newcommand{\Abt}{\tilde{\rm\bf A}}
\newcommand{\chit}{{\ts\tilde{\chi}}}

\newcommand{\dnb}{\partial_{n/b}}
\newcommand{\dnnb}{\partial_{n_0/b}}
\newcommand{\ohmnu}{{\ts\frac{1}{2}-\nu}}
\newcommand{\ohpnu}{{\ts\frac{1}{2}+\nu}}
\newcommand{\cpn}{\cos(\pi\nu)}
\newcommand{\rmrn}{(\rvec-\rvec_0)}
\newcommand{\delr}{{\scriptstyle\Delta}\rvec}

\begin{document}
\jl{1}

\title[Boundary integral method for magnetic billiards]%
{The boundary integral method for magnetic billiards}
\author{Klaus Hornberger\dag\ddag\ and Uzy Smilansky\dag}
\address{\dag\ Department of Physics of Complex Systems, 
The Weizmann Institute of Science, 76100 Rehovot, Israel}
\address{\ddag\ Max-Planck-Institut f\"ur Physik komplexer Systeme, 
N\"othnitzer Stra{\ss}e 38, 01187 Dresden, Germany}
\begin{abstract}
  We introduce a boundary integral method for two-dimensional quantum
  billiards subjected to a constant magnetic field. It allows to
  calculate spectra and wave functions, in particular at strong fields
  and semiclassical values of the magnetic length.  The method is
  presented for interior and exterior problems with general boundary
  conditions.
  We explain why the magnetic analogues of the field-free single and
  double layer equations exhibit an infinity of spurious solutions and
  how these can be eliminated at the expense of dealing with
  (hyper-)singular operators.  The high efficiency of the method is
  demonstrated by numerical calculations in the extreme semiclassical
  regime.
\end{abstract}
\vspace{\baselineskip}
\hspace{\fill}{\tt Preprint, 7 December 1999}

%
%
%
\section{Introduction}

Magnetic billiards are systems of a confined, charged particle in a
constant magnetic field.  In mesoscopic physics they serve as models
to explain shape-dependent features of nano-scale devices
\cite{NT88,URJ95}, like quantum dots. In quantum chaos they are
studied as natural extensions of planar billiards
\cite{RB85,BK96,Tasnadi97,Gutkin99}. These systems are particularly
suited for the study of semiclassical effects (both, theoretically
\cite{BB97,SNA98,Tanaka98} and in  experiments
\cite{MRWHG92,MWHG93,CBPW94}) since the field strength which essentially
determines the scale of quantum effects is a free parameter.

The presence of a Lorentz force severely affects the classical,
two-dimensional billiard dynamics. The criteria for hyperbolicity are
altered \cite{Tasnadi97,Gutkin99}. For strong enough fields closed
cyclotron orbits occur, while other trajectories perform a skipping
motion along the billiard boundary.  Most significantly, the exterior
dynamics where the billiard boundary acts as an obstacle from outside
is not a scattering problem like in the field free case but exhibits
bounded skipping motion around the billiard.

The magnetic quantum spectra and wave functions reflect these classical
properties. For strong fields a separation takes place in the
spectrum.  Close to the energies of the Landau levels one finds
\emph{bulk states} which correspond to a free cyclotron motion of the
particle. In addition, \emph{edge states} appear which are localized
at the boundary, corresponding to a skipping motion along it.  Unlike
the field free case, the spectrum is purely discrete also in the
exterior, with accumulation points at the energies of the Landau
levels.

From a technical point of view, calculations of spectra and wave
functions are considerably more difficult with a magnetic field
present.  So far, they were mostly realized by diagonalizing the
Hamiltonian \cite{BGOdAS95,YH95,ZLB95,RUJ96}.  This requires the
choice and truncation of a basis which is problematic in the general
case when no natural basis exists.  It explains why calculations of
exterior wave functions were not even attempted.

The spectra of field free billiards are usually calculated by
transforming the eigenvalue problem into an integral equation of lower
dimension. The corresponding integral operator is defined in terms of
the free Green function and depends only on the boundary.  This
\emph{boundary integral method} is known to be more efficient than
diagonalization by an order of magnitude and avoids the
arbitrariness of choosing and truncating a basis.

It seems natural to extend these ideas to the magnetic problem.  A
step in this direction was taken by Tiago \etal \cite{TCA97} who
essentially propose a null-field method \cite{Martin82} which involves
the irregular Green function in the angular momentum decomposition.  A
drawback of their approach is that the latter function must be known
for large angular momenta which is practically inaccessible
numerically.

In the present paper we propose a boundary integral method for
two-dimensional magnetic billiards. It involves the regular Green
function in the position space representation.
We derive the method for both, the interior and the exterior problem
and for general boundary conditions which include the Dirichlet and
Neumann choice as special cases.  The method allows for the first time
to calculate spectra and wave functions of magnetic billiards for
arbitrary fields and semiclassical values of the magnetic length.
Thousands of consecutive energy levels may be calculated to high
precision with moderate numerical effort.

%
\subsection{Outline}

For field-free billiards two independent boundary integral equations are
known. In section 2 we derive their magnetic analogues in a
gauge-invariant formulation.  It is shown that unlike in the
field-free case each of these equations yield only a necessary but not
a sufficient condition for the definition of the spectra. In other
words, each equation  admits spurious solutions. We will identify the
physical origin of the latter and propose a way to avoid them at the
expense of dealing with singular (and possibly even \emph{hypersingular}
\cite{Guiggiani98}) operators.

The explicit form of the integral operators is presented in section 3
where we discuss the nature of the singularities, too.  In section 4
it is shown how the integral equations may be solved treating the
singular parts of the operators analytically. This leaves the remaining
problem in a form suitable for numerical treatment. 
Its implementation is sketched in section \ref{sec:implementation}
together with a discussion of the numerical convergence and the
attainable accuracy.

Finally, we demonstrate the power of the proposed method in
section \ref{sec:results} where we study spectral statistics using
several thousand levels and present interior and exterior wave
functions in the quasi-classical regime.

%
%
\subsection{Preliminaries}

We are interested in the spectrum of a charged particle constrained to
a compact domain $\Domain \in \mathbb{R}^2$ which is subject to a
constant, perpendicular magnetic field of strength $B$.  Alternatively,
one may consider the complementary situation by constraining the
particle to the exterior $\mathbb{R}^2 \setminus
\openDomain$.  Unlike the non-magnetic case the
exterior spectrum is discrete, with accumulation points at the
Landau levels.
The stationary Schr\"odinger equation reads
\begin{equation}
  \frac{1}{2 m}
  \big( -\rmi \hbar {\grad}_{\!r} - q \mb{A}(\rvec) \big)^2
  \,  \psi(\rvec) 
  = E \, \psi(\rvec).
\end{equation}
where $m, q,$ and $E$ are mass, charge, and energy of the particle,
respectively.  The vector potential may be written in terms of the
symmetric gauge $\mb{A}^{\rm sym}$,
\begin{equation}
\label{eq:SG}
\mb{A}(\rvec)
=
\mb{A}^{\rm sym}(\rvec)  +  {\grad} \Chi(\rvec)
\defas
\frac{B}{2} \,r\, \mb{e}_\vartheta +  {\grad} \Chi(\rvec),
\end{equation}
where $\Chi$ accounts for the gauge freedom which we shall limit to
Coulomb type $\grad^{2} \Chi(\rvec)=0$.

We assume the domain boundary $\Gamma =\partial \Domain$ to be smooth
and choose its normals $\nvec(\rvec)$ to point outwards (i.e. into
$\mathbb{R}^2 \setminus \Domain$.)  Keeping their orientation fixed
will allow to distinguish the interior and the exterior problem.

The number of parameters ($E,B,\hbar,q,m$) in \eref{eq:SG} can be
reduced.  Scaling time by the Larmor frequency $\omega=q B/(2 m)$, one
remains with two length scales,
\begin{equation}
\rho^2=\frac{E}{2 m \omega^2} \qq\mbox{and}\qq b^2=\frac{\hbar}{m
\omega}, 
\end{equation}
as the only parameters describing the system.  $\rho$ is the classical
cyclotron radius whereas the magnetic length $b$ has a pure quantum
meaning. It gives the mean radius of a bulk ground state.  The scaled
energy may be expressed in terms of the spacing between Landau levels
\begin{equation}
\nu = \frac{E}{2 \hbar \omega} = \frac{\rho^2}{b^2}.
\end{equation}
The expression for the (unscaled) wave number $ k = {\sqrt{2 m
E}}/{\hbar} = {2 \rho}/{b^2} $ shows that there are two distinct
short-wave limits: The high-energy limit $\rho\to\infty$ and the
semiclassical limit $b\to 0$.  The former corresponds to increasing
the energy at fixed magnetic field while in the semiclassical limit
one increases both energy and field, keeping $\rho$ fixed.  It is
important to distinguish between these limits.  The high energy
direction is the simpler one because the dynamical effect of the
magnetic field tends to vanish.  However, for semiclassical studies
the later direction is the only proper choice because it leaves the
classical dynamics unaffected.

For most of the numerical demonstrations in section \ref{sec:results}
the magnetic length $b$ is chosen as the spectral variable in order to
present the boundary integral method in the nontrivial limit.
Therefore, to show clearly the dependence of the equations on $b$ we
do not introduce dimensionless variables for the scaled positions
$\rvec/b$. However, we facilitate the replacement by dimensionless
variables by stating all expressions (including the scaled gradient
$\grad_{\!r/b}\defas b\grad_{\!r}$) in terms of that quotient.

%
%
%
\section{Derivation of the boundary integral equations}

In this section two magnetic boundary integral equations are derived.
We show why they have spurious solutions and how to avoid this
by constructing a combined boundary integral equation.

\subsection{Single and double layer equations}

The quantum wave function $\psi\in\mathcal{L}_2(\mathbb{R}^2)$
is defined by its differential equation
\begin{eqnarray}
\label{eq:Schreq}
\left({\tfrac{1}{2}}[-\rmi\grad_{\!r/b} -
\tilde{\mb{A}}(\rvec)]^2 
-2\nu\right)
\psi(\rvec)
=0,
\end{eqnarray}
and a specification of boundary conditions.  We shall employ general
gauge invariant boundary conditions,
\begin{equation}
\label{eq:bcond}
\psi =  \pm\frac{\lambda}{b} (\dnb\psi-\rmi\At_n\psi),
\q\rvec\in\Gamma,
\end{equation}
with $\tilde{\mb{A}}=2/(Bb)\mb{A}(\rvec)$, $\At_n=\nvec(\rvec)\tilde{\mb{A}}$,
$\dnb\defas\nvec(\rvec)\grad_{\!r/b}\defas b\nvec(\rvec)\grad_{\!r}$.
Here, $\lambda$ (which may be a function of the position on the
boundary) interpolates between Dirichlet boundary conditions
($\lambda=0$) and the Neumann case ($\lambda^{-1}=0$).  
Equation \eref{eq:bcond} is a generalization of the mixed boundary
conditions known for the Helmholtz problem \cite{KSG64,BB70,SPSUS95}.  It
implies that the normal component of the current density operator
$\tilde{\jmath}_n={\rm Im}(\psi^*\dnb\psi)-\At_n|\psi|^2$ vanishes for
any $\lambda$.  The lower sign in \eref{eq:bcond} corresponds to the
exterior problem.

We mention in passing that another type of boundary conditions for
magnetic billiards was proposed recently \cite{AANS98}.

The Green function satisfies the inhomogeneous equation
\begin{eqnarray}
\label{eq:Gdef}
\left(\tfrac{1}{2}[-\rmi\grad_{\!r/b} - \tilde{\mb{A}}(\rvec)]^2 -2\nu\right)
{\rm G}(\rvec;\rvec_0)
= - \tfrac{1}{2}\,\delta\!\left(\tfrac{\rvec-\rvec_0}{b}\right).
\end{eqnarray}
Its properties are described in the appendix.  Note, that it does not
depend on the difference vector $\rmrn$ alone but has a gauge
dependent phase,
\begin{equation}
  \label{eq:GreenPhase}
  {\rm G}(\rvec;\rvec_0) = 
  \rme^{-\rmi\big( \tfrac{\rvec\times\rvec_0}{b^2}
                   -\chit(\rvec)+\chit(\mb{r_0})\big)}
  \,{\rm G}^0_\nu\Big(\tfrac{\rmrn^2}{b^2}\Big).
\end{equation}
We take ${\rm G}$ to be the free regular Green function by demanding
\begin{equation}
  \lim_{z\to\infty} {\rm G}^0_\nu(z) =0,
\end{equation}
which specifies ${\rm G}$ uniquely as the Fourier transform of the
free time evolution operator. As one expects, the regular Green
function decays exponentially once the points are separated by a
distance, $|\rmrn|>2\rho$, which cannot be traversed classically.  An
independent solution to \eref{eq:Gdef} exists which grows
exponentially beyond this classically allowed region. It may be called
irregular free Green function and was used in the null-field method
approach \cite{TCA97} for reasons to be explained below. In the
following only the regular Green function will be used.

We start by considering the \emph{interior} problem. The treatment of
the \emph{exterior} case is sketched afterwards.  Equations
\eref{eq:Schreq} and \eref{eq:Gdef} can be combined to yield a form
suitable for the Green and Gau{\ss} integral theorems,
\begin{equation}
  \label{eq:deleq}
  \psi\, \grad_{\!r/b}^2 \Gb 
  - \Gb\,\grad_{\!r/b}^2 \psi
  +2 \rmi\, \grad_{\!r/b} \big( \tilde{\mb{A}}\psi\Gb \big) 
  = \psi\,  \delta\!\left(\tfrac{\rvec-\rvec_0}{b}\right),
\end{equation}
where the bar indicates complex conjugation. 
Choosing $\rvec_0\in \mathbb{R}^2\setminus\Gamma$ the integral of
\eref{eq:deleq} over the domain $\Domain$ may be transformed to a
boundary integral,
\begin{equation}
\label{eq:unsplit}
\fl
\int_{\Gamma}
\big[
\psi\, \dnb\Gb-\Gb\,\dnb\psi+2\rmi\,\At_n\,\psi\,\Gb\;
\big] \frac{\rmd\Gamma}{b}
=
\Bigl\{
\begin{array}{cl}
\psi(\rvec_0)&\mbox{if $\rvec_0\in \openDomain $ }
\\
0&\mbox{if $\rvec_0\in \mathbb{R}^2 \setminus{\Domain}$, }
\end{array}
\end{equation}
corresponding to the interior problem.  Next, the vector potential
part of the integrand is split,
\begin{equation}
\label{eq:split}
\fl
\int_{\Gamma}
\big[
\psi\, (\dnb\Gb+\rmi\,\At_n\,\Gb)-\Gb\,(\dnb\psi-\rmi\,\At_n\,\psi)
\big] \frac{{\rm d}\Gamma}{b}
=
\Bigl\{
\begin{array}{cl}
\psi(\rvec_0)&\mbox{if $\rvec_0\in \openDomain $ }
\\
0&\mbox{if $\rvec_0\in \mathbb{R}^2 \setminus{\Domain}$. }
\end{array}
\end{equation}
which will allow for a gauge invariant formulation of the boundary
integral equation. Taking  $\rvec_0 \in
\Gamma$, $\rvec_0^\pm\defas\rvec_0\pm\epsilon \nvec_0$ with $\epsilon>0$,
we add the two equations above to obtain
\begin{equation}
\label{eq:SLepsin}
\int_{\Gamma}
\big[
\psi\, (\epscomb{\partial}_{n/b}\Gb +\rmi\, \At_n\, \epscomb{{\rm G}} )
- \epscomb{{\rm G}}\, (\dnb\psi-\rmi\,\At_n\,\psi) )
\big] 
\frac{{\rm d}\Gamma}{b}
=
{\tfrac{1}{2}}\psi(\rvec_0^-).
\end{equation}
Here we have introduced the abbreviations
$\epscomb{{\rm G}}={\tfrac{1}{2}}\Gb(\rvec;\rvec_0^+)
+{\tfrac{1}{2}}\Gb(\rvec;\rvec_0^-)$, 
$\epscomb{\partial}_{n/b}\Gb={\tfrac{1}{2}}\dnb\Gb(\rvec;\rvec_0^+)
+{\tfrac{1}{2}}\dnb\Gb(\rvec;\rvec_0^-)$.
Equation \eref{eq:SLepsin} is true for all (sufficiently small)
$\epsilon>0$ which means that the limit $\epsilon\to0$ exists.
Moreover, it can be shown that one is allowed to exchange the
integration with the limit ($\epscomb{{\rm G}}\to\Gb$,
$\epscomb{\partial}_{n/b}\Gb\to\partial_{n/b}\Gb$.)
Observing the boundary condition \eref{eq:bcond} and renaming the
unknown function, $u=\dnb\psi-\rmi\At_n\psi$, $u_0\defas u(\rvec_0)$
we get
\begin{equation}
  \label{eq:SingleLayerin}
\int_{\Gamma}
\big[
\,\Gb-\frac{\lambda}{b}
(\dnb\Gb +\rmi\, \At_n\, \Gb)
\big]\, u \,\frac{{\rm d}\Gamma}{b}
=
\frac{\lambda}{b}\, (- {\tfrac{1}{2}}u_0),
\end{equation}
which is an integral equation defined on the boundary $\Gamma$.

In order to obtain the corresponding equation for the exterior problem
consider a large disk $\mathcal{K}_p\supset\Domain$ of radius $p$ and
integrate \eref{eq:deleq} over $\mathcal{K}\cap\openDomain$.  Once
$\rvec_0$ is in the vicinity of $\Gamma$, the contribution of
$\partial\mathcal{K}$ to the boundary integral vanishes as
$p\to\infty$  due to the exponential decay
of ${\rm G}$ (since $\psi\in\mathcal{L}_2$).
Similar to \eref{eq:SLepsin} one  obtains an equation
\begin{equation}
\label{eq:SLepsout}
-
\int_{\Gamma}
\big[
\psi (\epscomb{\partial}_{n/b}\Gb +\rmi\, \At_n\, \epscomb{{\rm G}} )
- \epscomb{{\rm G}} (\dnb\psi-\rmi\,\At_n\,\psi) )
\big] 
\frac{{\rm d}\Gamma}{b}
=
{\tfrac{1}{2}}\psi(\rvec_0^+)
\end{equation}
which allows for the limit $\epsilon\to 0$ to be taken before
performing the integration.  The resulting boundary integral equation
differs from \eref{eq:SingleLayerin} only by a sign.  In the
following, we treat both cases simultaneously with the convention that
the upper sign stands for the interior problem and the lower sign for
the exterior case.
\begin{equation}
  \label{eq:SingleLayer}
\int_{\Gamma}
\big[\,
\Gb\mp\frac{\lambda}{b}
(\dnb\Gb +\rmi\, \At_n\, \Gb)
\big]\, u\, \frac{{\rm d}\Gamma}{b}
=
\frac{\lambda}{b}\, (- {\tfrac{1}{2}}u_0).
\end{equation}
For historical reasons  \cite{KR74}, we will refer to these
equations as the \emph{single layer equations} for the interior and
the exterior domain.

A second kind of boundary integral equations can be derived by applying
the differential operator $(\dnnb -\rmi
\At_{n_0})\defas\nvec(\rvec_0)(\grad_{\!r_0/b}-\rmi\tilde{\mb{A}}(\rvec_0)$
on equations \eref{eq:SLepsin} and \eref{eq:SLepsout},
\begin{eqnarray}
\label{eq:DLeps}
\fl
\int_{\Gamma}
 \psi\,
 ({\partial}_{n_0/b}-\rmi\, \At_{n_0})
 (\epscomb{\partial}_{n/b}\Gb+\rmi\, \At_n\,\epscomb{{\rm G}} ) 
\frac{{\rm d}\Gamma}{b}
- 
\int_{\Gamma}
 (\epscomb{\partial}_{n_0/b}\Gb -\rmi\, \At_{n_0}\, \epscomb{{\rm G}} )
 (\dnb\psi-\rmi\At_n\psi)
\frac{{\rm d}\Gamma}{b}
\nnn
=
\pm{\tfrac{1}{2}}
(\dnnb-\rmi\,\At_{n_0})\psi(\mb{r^\mp_0}).
\end{eqnarray}
This equation is true for all $\epsilon >0$ which means that the limit
$\epsilon\to 0$ exists. As for the first integral, we are again
allowed to permute the limit and the integration which yields a proper
integral.  Consequently, the limit of the second integral is finite,
too. However, in the second integral we are not allowed to exchange
the integration with taking the limit because the limiting integrand
has a $1/(\rvec-\mb{r_0})^2$-singularity which is not integrable.

Integral operators of this kind are named \emph{hypersingular}
\cite{Guiggiani98}. Similar to a Cauchy principal value integral,
they are defined by taking a special limit. However, in the present
case the singularity is stronger by one order. In the next section, we
define which limit is to be taken.  It is denoted by $\tfpint$ and
should be read ``finite part of the integral''. With this concept and
\eref{eq:bcond} we obtain the \emph{double~ layer} equations,
\begin{equation}
  \label{eq:DoubleLayer}
  \fl
  \int_{\Gamma}
   (\dnnb\Gb -\rmi\, \At_{n_0}\,\Gb  )
   \,u\,
   \frac{{\rm d}\Gamma}{b}
  \mp
  \frac{\lambda}{b}
   \fpint_{\Gamma}
   ({\partial}_{n_0/b} -\rmi\, \At_{n_0})
   ({\partial}_{n/b}\Gb +\rmi\, \At_{n}\,\Gb)
   \,u\, 
   \frac{{\rm d}\Gamma}{b}
  = 
  \mp \tfrac{1}{2}u
\end{equation}
which are again  integral equations defined on the boundary $\Gamma$.

It is useful to introduce a set of integral operators, (whose labels D
and N indicate correspondence to pure Dirichlet or Neumann conditions)
\begin{eqnarray}
  \label{eq:Opdef}
  \fl
  \mathsf{Q}^{\rm D}_{\rm sl} [u] = 
  \int_{\Gamma}\!
  {\rm d}\Gamma
  \,\Gb\, u,
\;\;&&
  \mathsf{Q}^{\rm N}_{\rm sl} [u] = 
  \int_{\Gamma}\!
  \frac{{\rm d}\Gamma}{b}
  (\dnb\Gb +\rmi\, \At_n\, \Gb)
  \,u,
\\
  \fl
  \mathsf{Q}^{\rm D}_{\rm dl} [u] = 
  \int_{\Gamma}\!
  \frac{{\rm d}\Gamma}{b}
   (\dnnb\Gb   -\rmi\, \At_{n_0}\,\Gb   )
   \,u,
\;\;&&
  \mathsf{Q}^{\rm N}_{\rm dl} [u] =  \,\,
   \fpint_{\Gamma}\!
   \frac{{\rm d}\Gamma}{b^2}
   ({\partial}_{n_0/b} -\rmi \At_{n_0})
   ({\partial}_{n/b}\Gb +\rmi\, \At_{n}\,\Gb)
    \,u.
   \nn
\end{eqnarray}
This way, the requirement of the existence of nontrivial solutions of
equations \eref{eq:SingleLayer} and \eref{eq:DoubleLayer}
is equivalent to demanding that the corresponding Fredholm 
determinants vanish,
\begin{eqnarray}
  \label{eq:detsl}
  \det\left[
    \mathsf{Q}^{\rm D}_{\rm sl}
    \mp \lambda
    \mathsf{Q}^{\rm N}_{\rm sl}
    +  \frac{\lambda}{2} \mathsf{id}
  \right] 
  &=& 0
  \q\q\mbox{(single layer,)}
\\
  \label{eq:detdl}
  \det\left[
    \mathsf{Q}^{\rm D}_{\rm dl}
    \mp \lambda
    \mathsf{Q}^{\rm N}_{\rm dl}
    \pm \frac{1}{2} \mathsf{id}
  \right] 
  &=& 0
  \q\q\mbox{(double layer.)}
\end{eqnarray}
These are secular equations although the explicit dependence on the
spectral variable is not shown in our abbreviated notation.  If one
chooses $\rho$ as the spectral variable, only the energy parameter
$\nu= \rho^2/b^2$ of the Green function is varied.  Taking $b$ as
spectral variable will in addition change the intrinsic length scale.

As mentioned already, each of the determinants \eref{eq:detsl}
and \eref{eq:detdl} may have zeros which do not correspond to
solutions of the original eigenvalue problem given by
\eref{eq:Schreq} and \eref{eq:bcond}.  For finite $\epsilon$ the
equations \eref{eq:SLepsin}, \eref{eq:SLepsout}, and \eref{eq:DLeps}
are still equivalent to the latter. They acquire additional spurious
solutions only as they are transformed to boundary integral equations
by the limit $\epsilon\to0$.

\subsection{Spurious solutions and the combined operator}

The physical origin of the redundant zeros is apparent in our gauge
invariant formulation.  They are proper solutions for the domain
\emph{complementary} to the one considered.  This is obvious for the
single layer equation with Dirichlet boundary conditions ($\lambda=0$)
where the spectral determinant does not depend on the orientation of
the normals. The same is true for the double layer equation with
Neumann boundary conditions ($\lambda^{-1}=0$).

In general, the character of the spurious solutions may be summarized
as follows: Independently of the boundary conditions, the
\emph{single layer} equation includes the \emph{Dirichlet} solutions
of that domain which is complementary to the one considered.
Likewise, the \emph{double layer} equation is polluted by the
\emph{Neumann} solutions of the complementary domain, irrespective of
the boundary conditions employed.

The statement is easily proved by observing that the
single-layer-Neumann operator and the double-layer-Dirichlet operator
are \emph{adjoint} to each other, \mbox{$\mathsf{Q}^{\rm N}_{\rm
sl}=(\mathsf{Q}^{\rm D}_{\rm dl})^\dagger$}, while the operators
$\mathsf{Q}^{\rm D}_{\rm sl}$ and $\mathsf{Q}^{\rm N}_{\rm dl}$ are
self-adjoint.  This is shown explicitly in the next section.  Now
assume that $u$ is a complementary Dirichlet solution. In Dirac notation,
\begin{eqnarray}
  \lo{}	
  \mathsf{Q}^{\rm D}_{\rm sl} \ket{u} =0
  \q\wedge\q
  \mathsf{Q}^{\rm D}_{\rm dl} \ket{u} \mp \tfrac{1}{2} \ket{u} =0
\\
  \lo{\Rightarrow}
  \bra{u}\mathsf{Q}^{\rm D}_{\rm sl} =0
  \q\wedge\q
  \bra{u}\mathsf{Q}^{\rm N}_{\rm sl} \mp \tfrac{1}{2} \bra{u} =0.
  \nn
\end{eqnarray}
Applying the dual of $u$ to the single layer operator yields
\begin{eqnarray}
  \bra{u} \mathsf{Q}^{\rm D}_{\rm sl}  
  \mp \lambda \Big\{\bra{u}\mathsf{Q}^{\rm N}_{\rm sl}  
                    \mp \tfrac{1}{2} \bra{u}\Big\}
  = 0,
\end{eqnarray}
which implies that the Fredholm determinant of the single layer
operator vanishes.  Similarly, if $u$ is a complementary Neumann
solution,
\begin{eqnarray}
  \lo{}	
  \pm\mathsf{Q}^{\rm N}_{\rm sl} \ket{u} + \tfrac{1}{2} \ket{u} =0
  \q\wedge\q
  \mathsf{Q}^{\rm N}_{\rm dl} \ket{u} =0
\\
  \lo{\Rightarrow}
  \pm\bra{u}\mathsf{Q}^{\rm D}_{\rm dl} + \tfrac{1}{2} \bra{u} =0
  \q\wedge\q
  \bra{u}\mathsf{Q}^{\rm N}_{\rm dl}  =0
  \nn
\end{eqnarray}
then its dual satisfies the double layer equation, again for any $\lambda$,
\begin{eqnarray}
  \pm\Big\{\pm\bra{u} \mathsf{Q}^{\rm D}_{\rm dl} + \tfrac{1}{2} \bra{u}\Big\}
  \mp \lambda \bra{u}\mathsf{Q}^{\rm N}_{\rm dl}
  = 0.
\end{eqnarray}
Since the spurious solutions are never of the same type it is possible
to dispose of them by requiring that both the single and the double
layer equations should be satisfied by the \emph{same} solution $u$.
Therefore, one obtains a necessary and sufficient condition for the
definition of the spectrum by considering a combined operator
\begin{equation}
  \label{eq:combdef}
  \mathsf{Q}_{\rm c}^\pm
  \defas
  \left(
    \mathsf{Q}^{\rm D}_{\rm dl}
    \mp \lambda
    \mathsf{Q}^{\rm N}_{\rm dl}
    \pm \frac{1}{2} \mathsf{id}
  \right)
  + \rmi \alpha    
  \left(
    \mathsf{Q}^{\rm D}_{\rm sl}
    \mp \lambda
    \mathsf{Q}^{\rm N}_{\rm sl}
    +  \frac{\lambda}{2} \mathsf{id}
  \right).
\end{equation}
with an arbitrary constant $\alpha$.  It has a zero eigenvalue only if
both, single and double layer operators do so.  The choice
\eref{eq:combdef} works very well in practice as will be shown below.
 
It seems natural to require that both the single and the double layer
equation must be satisfied to determine a proper eigenvalue. The
original equation \eref{eq:unsplit} consists of two independent
conditions ($\rvec_0 \in \openDomain$ and $\rvec_0\in \mathbb{R}^2
\setminus{\Domain}$.)  Only for special situations, such as the field
free problem, the two conditions are equivalent so that each of them
\emph{singly} provides the correct spectrum. For a discussion of the
field free case see e.g. \cite{EP95,THS97}.

It is interesting to note that (for the interior problem) spurious
solutions will not appear if one uses the irregular Green function.
The reason is that the gauge-independent part of this function is
\emph{complex} which destroys the mutual adjointness of the
operators.  This is why the irregular Green function had to be chosen
for the null-field method employed in \cite{TCA97}.  For the boundary
integral method the option to use this exponentially divergent
solution of \eref{eq:Gdef} is excluded since the corresponding
operator would get arbitrarily ill-conditioned once the size of the
boundary exceeds the cyclotron diameter.

Our last remark is concerned with the important case of Dirichlet
boundary conditions.  Here, one could as well derive a pair of
boundary integral equations that are \emph{not} gauge-invariant.
(Just set $\psi=0$ in \eref{eq:unsplit} and consider $u=\dnb\psi$.) Of
course, these equations would yield the proper gauge-invariant
eigen-energies of the problem. However, the energies of the additional
spurious solutions would depend on the chosen gauge and a
characterization of the latter in terms of solutions of a
complementary problem would not be possible.

\section{The boundary integral operators}

In this section we give explicit expressions for the boundary
integrals.  This allows to define the ``finite part integral'' appearing
in the double layer equation.

\subsection{Explicit expression for the integral kernels}
The form of the Green function \eref{eq:GreenPhase} leads to the
following expressions for the integral kernels $Q(\rvec;\rvec_0)$ of
the operators \eref{eq:Opdef},
$
  \big(\mathsf{Q}[u]\big)(\rvec_0) 
  = \int_\Gamma \rmd\Gamma Q(\rvec;\rvec_0) u(\rvec),
$
with $\nvec=\nvec(\rvec)$, $\nvec_0=\nvec(\rvec_0)$,
$\delr\defas\rmrn$, and $z\defas\delr^2/b^2$.
\begin{eqnarray}
\fl
\label{eq:QexpslD}
  Q_{\rm sl}^{\rm D}(\rvec;\rvec_0)
  =
  \rme^{\rmi\big( \tfrac{\rvec\times\rvec_0}{b^2}
                   -\chit(\rvec)+\chit(\mb{r_0})\big)}
  \,  \Gn_\nu(z)
\\
\label{eq:QexpslN}
\fl
  Q_{\rm sl}^{\rm N}(\rvec;\rvec_0)
  =
  \rme^{\rmi\big( \tfrac{\rvec\times\rvec_0}{b^2}
                   -\chit(\rvec)+\chit(\mb{r_0})\big)}
  \,
  \Big\{
    \rmi \frac{\delr\times\nvec}{b^2} 
    \, \Gn_\nu(z)
    + 2 \frac{\delr\,\nvec}{\delr^2} 
    \,
    z\frac{\rmd}{\rmd z}\Gn_\nu(z)
  \Big\},
\\
\label{eq:QexpdlD}
\fl
  Q_{\rm dl}^{\rm D}(\rvec;\rvec_0)
  =
  \rme^{\rmi\big( \tfrac{\rvec\times\rvec_0}{b^2}
                   -\chit(\rvec)+\chit(\mb{r_0})\big)}
  \,
  \Big\{
    \rmi \frac{\delr\times\mb{n_0}}{b^2} 
    \, \Gn_\nu(z)
    - 2 \frac{\delr\,\nvec_0}{\delr^2} 
    \,
    z\frac{\rmd}{\rmd z}\Gn_\nu(z)
  \Big\},
\\
\label{eq:QexpdlN}
\fl
  Q_{\rm dl}^{\rm N}(\rvec;\rvec_0)
  =
  \rme^{\rmi\big( \tfrac{\rvec\times\rvec_0}{b^2}
                   -\chit(\rvec)+\chit(\mb{r_0})\big)}
  \,
  \Big\{
    (-\frac{(\delr\times\nvec_0)(\delr\times\nvec)}{b^4}
      +\rmi \frac{\nvec\times\nvec_0}{b^2}
      )   \Gn_\nu(z)
\\
    + ( 2\rmi \frac{\nvec\times\nvec_0}{b^2}
        - 2  \frac{\nvec\,\nvec_0}{\delr^2}       )    
       \,
       z\frac{\rmd}{\rmd z}\Gn_\nu(z)  
    - 4   \frac{(\delr\nvec)(\delr\nvec_0)}{\delr^4}
       \,       
       z^2\frac{\rmd^2}{\rmd z^2}\Gn_\nu(z)  
  \Big\}.
\nn
\end{eqnarray}
Note, that the gauge freedom $\Chi$ cancels in the prefactors and only
appears in the phase. It can be absorbed by the function $u(\rvec) \to
\exp(-\rmi\Chi(\rvec)) u(\rvec)$ proving the manifest gauge invariance
of the boundary integral equations \eref{eq:SingleLayer},
\eref{eq:DoubleLayer}.  It can also be seen easily that expressions
\eref{eq:QexpslN} and \eref{eq:QexpdlD} are related by a permutation
of $\rvec$ and $\rvec_0$ with subsequent complex conjugation (since
$\Gn_\nu$ is real), i.e. the operators are the adjoints of each
other. The self-adjoint nature of \eref{eq:QexpslD} and
\eref{eq:QexpdlN} follows likewise.

The gauge independent part of the Green function, $\Gn_\nu$, has a
logarithmic singularity at $\rvec=\rvec_0$. Its derivatives appearing
in \eref{eq:QexpslN} -- \eref{eq:QexpdlN} can be expressed in terms of
$\Gn_\nu$ itself, at different energies $\nu$, and may be found in the
appendix \eref{eq:zdG}, \eref{eq:zzddG}.
They are bounded as $\rvec\to\rvec_0$.
In that limit most of the quotients vanish for a smooth boundary,
others tend to the curvature $\kappa_0$ at the boundary point $\rvec_0$
(defined to be positive for convex domains,)
\begin{equation}
  \label{eq:curvlim}
  \lim_{\rvec\to\rvec_0}\frac{\rmrn\nvec}{\rmrn^2}=\frac{\kappa_0}{2},
  \qq
  \lim_{\rvec\to\rvec_0}\frac{\rmrn\mb{n_0}}{\rmrn^2}=-\frac{\kappa_0}{2}.
\end{equation}
As a consequence, all the terms in \eref{eq:QexpslD} -- \eref{eq:QexpdlN}
are integrable but for the one containing the
$(\nvec\,\nvec_0)/\delr^2$-singularity.  The latter gives rise to
the need for a finite part integral.

\subsection{The hypersingular integral operator}
For finite $\lambda$ the double-layer equation contains a
hypersingular integral defined as
\begin{eqnarray}
\label{eq:fpi}
\fl
  \mathsf{Q}^{\rm N}_{\rm dl} [u] 
  \,\,=\,\, 
  \fpint_{\Gamma}
   \frac{{\rm d}\Gamma}{b^2}
   ({\partial}_{n_0/b} -\rmi \At_{n_0})
   ({\partial}_{n/b}\Gb +\rmi \At_{n}\Gb)
   u 
\nnn
  \defas 
  \lim_{\epsilon\to0}
  \int_{\Gamma}
   \frac{{\rm d}\Gamma}{b^2}
   ({\partial}_{n_0/b} -\rmi \At_{n_0})
   (\epscomb{\partial}_{n/b}\Gb +\rmi \At_{n}\epscomb{\rm G})
    u.
\end{eqnarray}
We want to replace the integrand by its
limiting form.  To this end the boundary is split into the part
$\gamma_{c\epsilon}$ within an $(c\epsilon)$-vicinity around $\rvec_0$
(with arbitrary constant $c$) and the remaining part
$\Gamma_{\!c\epsilon}$,
\begin{eqnarray}
  \label{eq:fpieps}
 =  \lim_{\epsilon\to0}
  &\bigg[&
  \int_{\Gamma_{\!\scriptstyle c\epsilon}}
    \!\! \frac{{\rm d}\Gamma}{b^2}
    (\partial_{n_0/b} -\rmi \At_{n_0})
    (\epscomb{\partial}_{n/b}\Gb +\rmi \At_{n}\epscomb{\rm G})
     u
\nnn
   &+&  \int_{\gamma_{\scriptstyle c\epsilon}}
    \!\! \frac{{\rm d}\Gamma}{b^2}
    (\partial_{n_0/b} -\rmi \At_{n_0})
    (\epscomb{\partial}_{n/b}\Gb +\rmi \At_{n}\epscomb{\rm G})
     (u-u_0)
\\
    &+&
    u_0
  \int_{\gamma_{\scriptstyle c\epsilon}}
    \!\! \frac{{\rm d}\Gamma}{b^2}
    (\partial_{n_0/b} -\rmi \At_{n_0})
    (\epscomb{\partial}_{n/b}\Gb +\rmi \At_{n}\epscomb{\rm G})
  \bigg],
\nn
\end{eqnarray}
with $u_0 \defas u(\rvec_0)$.  For sufficiently small $\epsilon$ the
boundary piece $\gamma_{c\epsilon}$ may be replaced by its tangent and
the Green function by its asymptotic expression, cf. appendix A.  This
way the third integral in \eref{eq:fpieps} may be evaluated to its
contributing order,
\begin{eqnarray}
  \label{eq:fpixxx}
\fl
  \int_{\gamma_{\scriptstyle c\epsilon}}
    \!\! \frac{{\rm d}\Gamma}{b^2}
    (\partial_{n_0/b} -\rmi \At_{n_0})
    (\epscomb{\partial}_{n/b}\Gb +\rmi \At_{n}\epscomb{\rm G})
\\
\fl
  =
  \frac{1}{4\pi}
  \int_{-c\epsilon}^{c\epsilon}
    \!\!\!    \!\!\!
    \rmd s
    \cos\! \Big(\frac{\rvec_0\nvec_0}{b^2}s\Big)
    \cos\! \Big(\epsilon(\frac{\nvec_0\times\rvec_0}{b^2}-s)\Big)
    \bigg[
      \frac{-2}{s^2+\epsilon^2}
      +4\frac{\epsilon^2}{(s^2+\epsilon^2)^2}
    \bigg]
    \,\,
  + \Or(\epsilon^2\log\epsilon)  
\nnn
\fl
 =
  \frac{1}{2\pi}
  \int_{-c\epsilon}^{c\epsilon}
    \!\!\!    \!\!\!
    \rmd s
      \frac{\epsilon^2-s^2}{(s^2+\epsilon^2)^2}
    \,\,
  + \Or(\epsilon^2\log\epsilon)  
 =
  \frac{1}{\pi}   \frac{1}{c\epsilon}
  \frac{c^2}{c^2+1}
  + \Or(\epsilon^2\log\epsilon)
 \approx
  \frac{1}{\pi}   \frac{1}{c\epsilon}
  + \Or(\epsilon^2\log\epsilon).  
  \nn
\end{eqnarray}
Here, the explicit form of the integrand was obtained from
\eref{eq:QexpdlN} by the replacement $\rvec_0\to\rvec_0^\pm$.
The last approximation in \eref{eq:fpixxx} holds because $c$ may be
chosen arbitrarily large.  In a similar fashion it can be shown that
the second integral in \eref{eq:fpieps} is of order
$\Or(\epsilon)$. In the first integral we may replace (again for large
$c$) the integrand by its limit because $\epsilon$ is small compared
to $\min(|\rvec-\rvec_0|)=c\;\epsilon$.  Therefore, the limit in
\eref{eq:fpi} may be expressed as
\begin{eqnarray}
\label{eq:fpilim}
  \fpint_{\Gamma}
   \frac{{\rm d}\Gamma}{b^2}
   ({\partial}_{n_0/b} -\rmi \At_{n_0})
   ({\partial}_{n/b} +\rmi \At_{n})
   \Gb u 
\nnn
  =  
 \lim_{\epsilon\to0}
 \bigg[
  \int_{\Gamma_{\!\scriptstyle \epsilon}}
    \!\! \frac{{\rm d}\Gamma}{b^2}
    (\partial_{n_0/b} -\rmi \At_{n_0})
    ({\partial}_{n/b}\Gb +\rmi \At_{n}\Gb)
     u
   \,\,
   + u_0 \frac{1}{\pi\epsilon}
 \bigg],
\end{eqnarray}
where we replaced $c\epsilon$ by $\epsilon$. This equation
defines the finite part integral.  It completes the derivation
of the boundary integral equations and we may now turn to the question
of how to solve them.

%
%
%

\section{Solving the integral equations}
\label{sec:solv}

As shown above, the integral equations \eref{eq:SingleLayer} and
\eref{eq:DoubleLayer} may be used to compute spectra of magnetic
billiards.  However, the corresponding integral kernels are not yet
suitable for numerical evaluation.  In this section we show how their
asymptotically singular behaviour may be separated and be treated
analytically.

In the following the \emph{combined} integral equation as defined by
\eref{eq:combdef} will be considered.  The corresponding expressions
for the pure double layer or single layer case may be obtained easily
by setting $\alpha=0$ or $\alpha^{-1}=0$, respectively.
We also take the opportunity to \emph{regularize} the integral
equations.  At the energies of the Landau levels,
$\nu_n=n+\tfrac{1}{2}, n\in\mathbb{N}_0$, they are defined only by the
limit $\nu\to\nu_n$, so far.  This is because the Green function is
singular at the energies $\nu_n$.  These simple poles are removed by
multiplying the equations with $\cpn$ and taking the limiting values
at $\nu_n$.

For convenience we assume $\lambda$ to be constant on $\Gamma$ and the
domain $\Domain$ to be simply connected. Let its boundary of length
$\Len=|\Gamma|$ be parameterized by the arc length $s$,
\begin{equation}
  \label{eq:parametr}
  \Gamma: s \in [0;\Len] \mapsto \rvec(s) 
  \q\mbox{with}\q
  \frac{{\rm d}\rvec(s)}{{\rm d}s} 
  \defas \tvec(s) 
  = {-n_y(s)\choose n_x(s)},
\end{equation}
which allows to write the (regularized) integral kernel 
\begin{eqnarray}
\label{eq:defq}
  {\rm q}(s,s_0) \defas
  \cpn \big[ Q^{\rm D}_{\rm dl}(\rvec_s;\rvec_{s_0}) 
            +\rmi\alpha Q^{\rm D}_{\rm sl}(\rvec_s;\rvec_{s_0}) 
\nnn
 \qq\qq\qq\q  \mp \lambda 
             \big(
                Q^{\rm N}_{\rm dl}(\rvec_s;\rvec_{s_0}) 
                +\rmi\alpha Q^{\rm N}_{\rm sl}(\rvec_s;\rvec_{s_0}) 
             \big)
       \big]
\end{eqnarray}
with $\rvec_s\defas\rvec(s)$. After an expansion of the boundary
around $\rvec(s_0)$,
\begin{equation}
  \rvec(s) = 
  \rvec_0 +  (s-s_0)\,\tvec_0 - \frac{\kappa_0}{2}(s-s_0)^2\, \nvec_0 
  + \Or\Big((s-s_0)^3\Big),
\end{equation}
one obtains, observing \eref{eq:QexpslD} --- \eref{eq:QexpdlN}, the
asymptotic behaviour for small $s'=s-s_0$,
\begin{eqnarray}
  \label{eq:asympq} 
\fl
  {\rm q}(s_0+s',s_0) \defas
  \rme^{\rmi\tfrac{\rvec_s\times\rvec_0}{b^2}} 
  \,
  \bigg\{
     \mp\lambda\,\frac{\cpn}{2\pi}
     \, \frac{-1}{{s'}^2} 
\nnn
     +\Big[ -\rmi\frac{s'}{b^2}+\rmi\alpha
      \mp\lambda\Big(
		\frac{2\nu}{b^2} + (\alpha-\rmi\kappa_0) \frac{s'}{b^2}
                \Big)
      \Big] 
      \,{\rm L}_\nu\Big(\frac{{s'}^2}{b^2}\Big)
\nnn
     +\Big[\kappa_0\mp\lambda\Big(-2\frac{\nu}{b^2}
                                           +\rmi\alpha\kappa_0\Big)\Big]
      \, \frac{\cpn}{4\pi}
      \,+\, \Or({s'}^2\log {s'}^2)
  \bigg\}.
\end{eqnarray}
The necessary asymptotic expansions for the gauge-independent part of the Green
function and its derivatives may be found in the appendix.  ${\rm
L}_\nu$ describes the asymptotically logarithmic form of the Green
function and is defined in \eref{eq:defL}.  Note that due to the
quotient $1/{s'}^2$ the expansion of $z\partial_z \Gn$ contributes up to
and including order $\Or({s'}^2\log {s'}^2)$. Similarly, the second order
term of $\nvec\nvec_0 = 1-\tfrac{1}{2}\kappa_0^2 {s'}^2+\Or({s'}^3)$ 
enters with the effect of cancelling another term.

\begin{figure}
\epsfxsize\textwidth%
\epsfbox{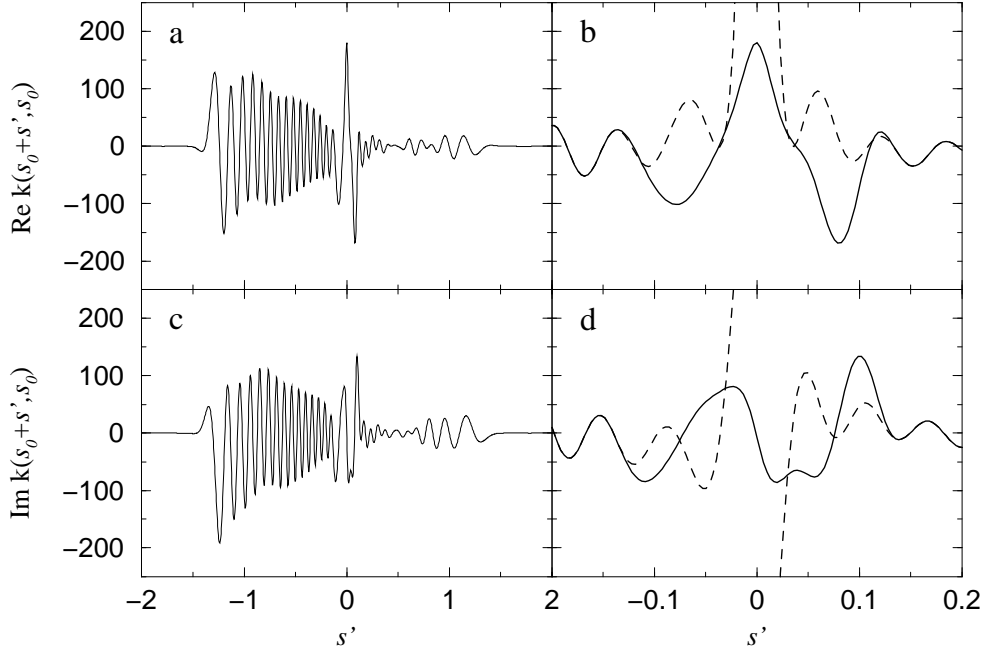}
\caption{%
(a) Real and (c) imaginary part of the smooth combined integral kernel
\eref{eq:defk} for fixed $s_0$ and the case of Neumann boundary
conditions. We choose $\rho=0.6$ and an elliptic domain (of
eccentricity $0.8$ and area $\Area=\pi$, centered on $(0.5,0.25)$) at
$\nu=19$ corresponding to the energy of the $\sim 1000^{\rm th}$
interior eigenstate.  The boundary point $s_0=0$ is  that
of largest curvature.  The magnifications (b) and (d) around $s'=0$
include the original singular kernel \eref{eq:defq} as a dashed line.
}
\label{fig:smoothkernel}
\end{figure}

As apparent from \eref{eq:asympq}, the singularities of the integral
kernel are well described by the functions
\begin{equation}
\label{eq:defm}
  {\rm m}(s,s_0) \defas
  \mp\lambda
  \, \rme^{\rmi\tfrac{\tvec_0\times\rvec_0}{b^2}(s-s_0)} 
  \, \frac{\cpn}{2\pi}
  \frac{-1}{(s-s_0)^2}
\end{equation}
and, for the logarithmic part,
\begin{eqnarray}
\label{eq:defl}
  {\rm l}(s,s_0) \defas
  \rme^{\rmi\tfrac{\tvec_0\times\rvec_0}{b^2}(s-s_0)} 
 \,{\rm L}_\nu\Big(\frac{(s-s_0)^2}{b^2}\Big)
  \nnn
  \qq \times
    \Big[ \rmi\alpha -\rmi\frac{(s-s_0)}{b^2}
      \mp\lambda\Big(
                 \frac{2\nu}{b^2}
		 +(\alpha-\rmi\kappa_0) \frac{(s-s_0)}{b^2}
                 \Big)
      \Big].
\end{eqnarray}
It is important to include the terms of order
$\Or\big(s\log(s^2)\big)$ to ensure that the smooth integral kernel
defined as
\begin{eqnarray}
\label{eq:defk}
  {\rm k}(s,s_0) \defas
    {\rm q}(s,s_0) - g(s-s_0)\,
             \big[     {\rm l}(s,s_0) + {\rm m}(s,s_0)
             \big] 
\end{eqnarray}
is differentiable at $s=s_0$ (provided the curvature is continuous).
Here, $g(s')$ is a window function (with $g(0)=1$) which
smoothly switches off the singular functions for $|s'|>0$ and vanishes
beyond some small, suitably  chosen $\srange$.  Figure
\ref{fig:smoothkernel} shows the smooth as well as the original kernel
for a typical choice of the boundary and the energy.

%
%

The solution $u(s)$ of the boundary integral equation is periodic and
may therefore be expanded in a Fourier series,
\begin{equation}
  \label{eq:fourieru}
  u(s)\,    \rme^{-\rmi\chit(s)}
   = 
   \sum_{\ell=-\infty}^\infty u_\ell 
   \, \rme^{\ts 2\pi\rmi \ell s/\Len}.
\end{equation}
As mentioned above, we include the phase due to the gauge freedom
$\chit$ which amounts to the choice of the symmetric  gauge for the
actual calculation.  Within the Fourier representation 
the Fredholm determinant may be written in the form
\begin{equation}
\label{eq:Fourierdet}
  \det\Big[
          {\rm K}_{k\ell}+{\rm L}_{k\ell}+{\rm M}_{k\ell}
          -\Len\, c\, \delta_{k\ell}
  \Big]_{k,\ell\in\mathbb{Z}} = 0
\end{equation}
with $c\defas (\mp\tfrac{1}{2}-\rmi\tfrac{1}{2}\alpha\lambda)\cpn$.
It consists of a double Fourier integral over
the smooth kernel,
\begin{eqnarray}
  \label{eq:defKkl}
   {\rm K}_{k\ell} \defas
      \int\!\!\!\!\int_{\Len^2} 
      \!\!\!\!\rmd s_0 \rmd s
      \,\,
      \rme^{\ts 2\pi\rmi (s\ell-s_0k)/\Len}
      \;
      {\rm k}(s,s_0)
\end{eqnarray}
and two single  Fourier integrals,
\begin{eqnarray}
  \label{eq:defLkl}
   {\rm L}_{k\ell} 
   \defas
   \int_{\Len}
     \!\!  \rmd s_0 
     \,\, \rme^{\ts 2\pi\rmi s_0 (\ell-k)/\Len}
     \; {\rm L}_\ell(s_0)
\q\mbox{and}
\\
  \label{eq:defMkl}
   {\rm M}_{k\ell} 
   \defas
   \int_{\Len}
     \!\!  \rmd s_0 
     \,\, \rme^{\ts 2\pi\rmi s_0 (\ell-k)/\Len}
     \; {\rm M}_\ell(s_0).
\end{eqnarray}
Here, ${\rm L}_\ell(s_0)$ and ${\rm M}_\ell(s_0)$ are (finite part)
Fourier integrals over the asymptotic singularities,
\begin{eqnarray}
  \label{eq:defLl}
   {\rm L}_\ell(s_0) = 
   \int_{-\srange}^{\srange}
   \!\!  \rmd s' 
   \,\, \rme^{\ts 2\pi\rmi \ell s'/\Len}
   \;
   g(s')
   \, {\rm l}(s_0+s';s_0)
\q\mbox{and}
\\
  \label{eq:defMl}
   {\rm M}_\ell(s_0) = 
   \fpint_{-\srange}^{\srange}
   \!\!  \rmd s' 
   \,\, \rme^{\ts 2\pi\rmi \ell s' /\Len}
   \;
   g(s')
   \, {\rm m}(s_0+s',s_0).
\end{eqnarray}
They may be calculated \emph{analytically}, for a suitable window $g$,
yielding smooth functions of $s_0$.  In appendix B the results can be
found for
\begin{equation}
  \label{eq:defg}
  g(s') \defas 
  \cos^{2}\!\!\Big(\frac{\pi}{2}\frac{s'}{\srange}\Big)
  \Big(   \Theta(s'-\srange)-\Theta(s'+\srange)
  \Big),     
\end{equation}
where $\Theta$ is the Heaviside step function.  With this choice of
the window function they are given in terms of elementary functions
and may be evaluated easily.
Having treated the (hyper-)singular features of the boundary integrals
analytically, the remaining problem can be solved efficiently by
numerical means.

%
%
%
%
%
\section{Numerical Analysis}
\label{sec:implementation}

In the following, we describe shortly some aspects of the numerical
treatment and discuss the question of numerical accuracy. 

The evaluation of the remaining Fourier integrals \eref{eq:defKkl} -
\eref{eq:defMkl} must be performed numerically.  Since the integrands
are well-behaved this may be done  by dividing the boundary into
$N$ equidistant pieces and approximating the integrand at each one by
its value at the mid-point.  The summations may be performed by a
Fast-Fourier algorithm.  For large enough $N$ this simple method is more
effective than any attempt to evaluate the highly oscillatory
integrals \eref{eq:defKkl} -- \eref{eq:defMkl} by more sophisticated
schemes.

%
%
\begin{figure}
\epsfxsize\textwidth%
\epsfbox{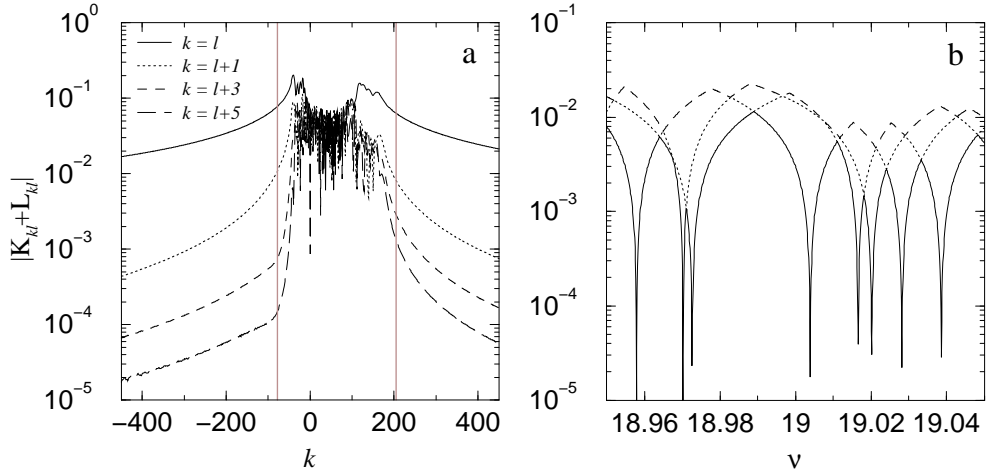}
\caption{%
(a) Matrix ${\rm K}_{k\ell}+{\rm L}_{k\ell}$ corresponding to the same
choice of parameters as in Fig.~\ref{fig:smoothkernel} and Dirichlet
boundary conditions. Shown are the absolute values of the matrix
elements along its diagonal and neighboring diagonals. Apart from the
diagonal, appreciable values of the matrix ${\rm K}_{k\ell}+{\rm
L}_{k\ell} -\Len\, c\, \delta_{k\ell}$ are localized within a
sub-block which allows safe truncation. The vertical lines indicate
the typical size after truncation.  (b) The three smallest singular
values of the matrix around $\nu=19$ (at constant $\rho=0.6$
corresponding to roughly the $1000^{\rm th}$ eigenvalue.) The minima
of the smallest singular value (solid line) determine the spectrum to
a high accuracy.  }
\label{fig:MandS}
\end{figure}

Due to the Fourier representation the resulting large $N\times N$
matrix has a partly diagonal structure, cf. Fig.  \ref{fig:MandS}(a).
There are off-diagonal elements of appreciable value only within a
sub-block the size of which is independent of $N$.  Outside of the
sub-block essentially only the diagonal elements are occupied (the
values decay rapidly as one leaves the diagonal.)  It is the
\emph{bulk} wave functions which are given by the null vectors
corresponding to the latter diagonal Fourier components.
These components do not contribute to the other states since they are
not coupled to them. As a consequence, the restriction of the matrix
to the above-mentioned sub-block at most removes bulk states, if they
exist, out of the numerically calculated spectrum \emph{without}
affecting other states.  Generically, one is not particularly
interested in these states whose energies are exponentially close to
the Landau levels.  Since the spectrum is modified at most in a
well-controlled way, it is permissible to truncate the matrix to a
smaller size $N_{\rm trunc}$.

A small complication arises in the case of finite $\lambda$.  Due to
the hypersingular part the diagonal Fourier elements increase linearly
as $|\ell|\to\infty$, cf.~\eref{eq:Masymp}.  The above statements
apply in this case after dividing the matrix \eref{eq:Fourierdet}
column-wise by the asymptotic factor
\begin{eqnarray}
	\Big[
	\big(\frac{<\!\tvec_0\times\rvec_0\!>}{b^2} 
          + 2\pi \frac{\ell}{\Len}\big)^2
	+\big(\frac{{\rm Si}(\pi)}{\srange}\big)^2
	\Big]^{1/2}_.
\end{eqnarray}
Here, $<\!\tvec_0\times\rvec_0\!>$ is the average (the $0^{\rm th}$
Fourier component) of the function $\tvec(s_0)\times\rvec(s_0)$
defined on the boundary.  

The calculation of the spectrum amounts to finding (all) the zeros of
the complex-valued determinant \eref{eq:Fourierdet} in a given energy
range.  Numerically, this is the most expensive task, scaling as
$N_{\rm trunc}^3$.  Since the computation of the determinant tends to
be unstable around its zeros it is more advantageous to employ a
singular-value decomposition of the matrix which is stable in any
case. The vanishing of a singular value indicates a defective rank of
its matrix. Due to roundoff errors these non-negative quantities are
always greater than zero. However, the spectral points are very well
defined by the sharp minima of the lowest singular value as a function
of $\nu$, cf.  Figure \ref{fig:MandS}(b).  The detection of near
degeneracies is made appreciably easier if one monitors the next
smallest singular values, too.

In order to calculate the wave function
$\psi_0=\psi(\rvec_0\notin\Gamma)$ away from the boundary one may use
directly equation \eref{eq:split}.  The gauge invariant gradient of
the wave function,
$\boldsymbol{\gamma}_0\defas\grad_{r/b}\psi_0-\rmi\Abt_0\psi_0$ needed
for the current density $\boldsymbol{\tilde{\jmath}}_0={\rm
Im}[\psi^*_0\boldsymbol{\gamma}_0]$ is obtained from the same equation
after the application of the operator $\grad_{r_0/b}-\rmi\Abt_0$.
\begin{eqnarray}
\fl
\psi_0
&=&
\!\pm\!
\int_{\Gamma}
\!
\frac{{\rm d}\Gamma}{b}
\big[
\pm\tfrac{\lambda}{b} (\dnb\Gb+\rmi\At_n\Gb)-\Gb
\big] u,
\\
\fl
\boldsymbol{\gamma}_0
&=&
\!\pm\!
\int_{\Gamma}
\!
\frac{{\rm d}\Gamma}{b}
\big[
    \pm\tfrac{\lambda}{b} (\grad_{r_0/b}-\rmi\Abt_0)(\dnb\Gb+\rmi\At_n\Gb)
    -(\grad_{r_0/b}\Gb-\rmi\Abt_0\Gb)
\big] u 
\end{eqnarray}
Since the integrands are not singular for $\rvec_0\notin\Gamma$ the
integrals may be approximated by a discrete sum over the $N$ boundary
elements without further ado.  The densities of other observables can
be obtained by similar boundary integrals.

%
%
\begin{figure}
\epsfxsize\textwidth%
\epsfbox{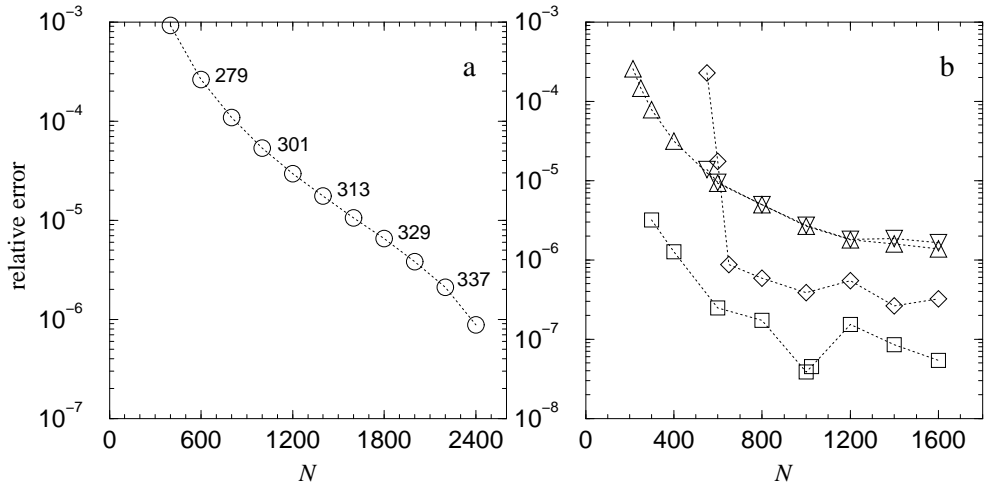} 
\caption{%
Errors of the $\sim 1000^{\rm th}$ interior eigenvalue at $\rho=0.6$
as a function of the boundary discretization $N$.  (a) Approximate
relative error for the elliptic domain of Fig.  \ref{fig:MandS}(b)
(the Dirichlet state closest to $\nu=19$) Here, the energy for
$N=2600$ was taken as reference. The numbers indicate the matrix
dimension after truncation which determines the numerical effort. They
increase only weakly with $N$.  (b) Exact relative error of the
\emph{exterior Neumann} energies of a typical edge state (
$\opentriangle, \opentriangledown$) and a typical bulk state
($\opensquare, \opendiamond$) as a function of $N$. Here, we use a
\emph{circular domain} (of area $\Area=\pi$) which allows to determine
the exact energies ($\nu_{\rm edge}\simeq19.0294509$, $\nu_{\rm
bulk}\simeq19.4816851$) independently.  The center of the domain is
placed at the origin ($\opentriangle, \opensquare$) and at the point
(3,0) ($\opentriangledown, \opendiamond$), respectively.  One observes
that the displacement does not affect the error of the edge state but
increases the error of the bulk state energy systematically.  [Note,
that the graphs do not have the same scale.]  }
\label{fig:accuracy}
\end{figure}

\subsection{Convergence and Accuracy}

Careful numerical tests indicate that the precision of the calculated
spectra and wave functions is determined almost exclusively by the
dimension $N$ of the initial matrix.  In Figure \ref{fig:accuracy}(a)
we show how the energies converge exponentially as $N$ increases.  At
the same time, the calculated spectra are found to be numerically
invariant with respect to other parameters such as $\alpha$,
$\srange$, $N_{\rm trunc}$, and in particular the location of the
origin.

Reasonable choices for $\alpha$ and $\srange$ are $\alpha=\nu/(2\rho)$
and $\srange=b$.  The location and size of the sub-block is best
determined in terms of an averaged column norm. The resulting spectra
are independent of $N_{\rm trunc}$ provided it exceeds a critical
value.  Here, the position of the origin is relevant, because the
calculation of the spectral determinant \eref{eq:Fourierdet}, in
particular its analytical parts, must be performed in a specific
gauge.  The choice in favor of the symmetric gauge is made in
\eref{eq:fourieru} where the remaining gauge freedom $\chi$ is
absorbed into the solution vector.  As a consequence of the resulting
distinction of the origin, the spectral determinant is no longer
translationally invariant.

As a result, the size of the truncated matrix depends on the choice of
the origin. For example, the values in Fig. \ref{fig:accuracy}(a)
belong to an ellipse centered at the origin. With an ellipse displaced
by (2,1) one obtains the {same} relative errors for $N=600 \ldots
2400$ (not shown, one would not see a difference) with truncation
sizes larger by 50\%.  In order to minimize the numerical effort it is
therefore advantageous to choose the origin in the center of the
domain considered.

The fact that bulk states are more strongly affected by the truncation
is seen in Fig.~\ref{fig:accuracy}(b) where exterior Neumann states of
a circular domain are compared after displacement by 3 radii.  Since
the disk is a separable problem, we can check here against the exact
energies (obtained as the roots of an analytical function.)  Note,
that the calculation of the hypersingular integral introduces no
additional error.

The only precise published calculations for a nontrivial shape known
to us are the results of Tiago \etal who give the first twenty Dirichlet
levels for an ellipse of eccentricity $0.8$ and area $\Area=\pi$ at constant
$b^2=2/25$ (missing one symmetry class.) Our method is able to confirm
their results to \emph{all} given seven digits (apart from occasional
differences in the last digit by one).
For reference, we note the energy of the approximately
\emph{thousandth}
state (the one closest to $\nu=80$) which we calculate to be
$\nu\simeq79.9362(6)$. The expected error is less than $0.1\%$ of the
mean level spacing.

%
%
\begin{figure}
\epsfxsize\textwidth%
\epsfbox{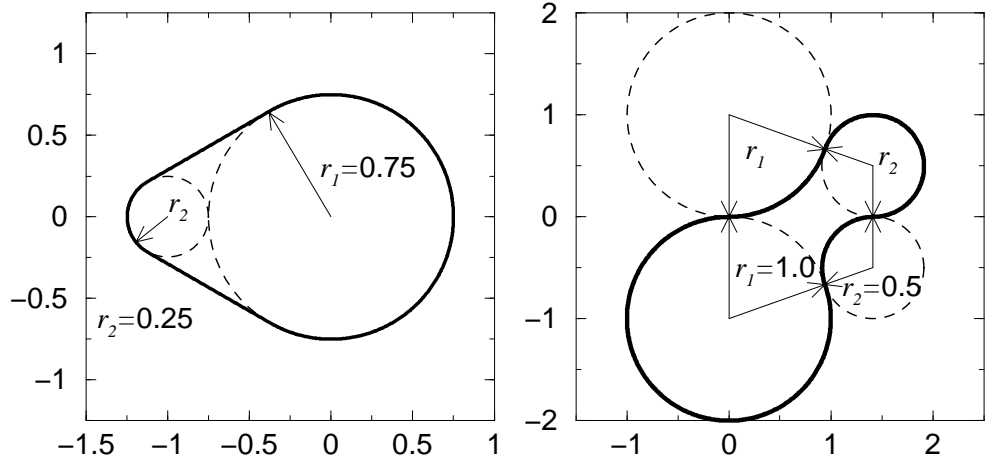} 
\caption{%
Definition of the domain boundaries considered in section
\ref{sec:results}.  In the asymmetric stadium (left) the magnetic
dynamics shows no unitary but one anti-unitary symmetry. In contrast, the
skittle shaped domain (right) is free of any symmetry. It generates
\emph{hyperbolic} classical motion even for strong magnetic fields
$\rho>1$. }
\label{fig:shapes}
\end{figure}

%
%
%
%
\section{Numerical Results}
\label{sec:results}

%
%
\begin{figure}
\epsfxsize\textwidth%
\epsfbox{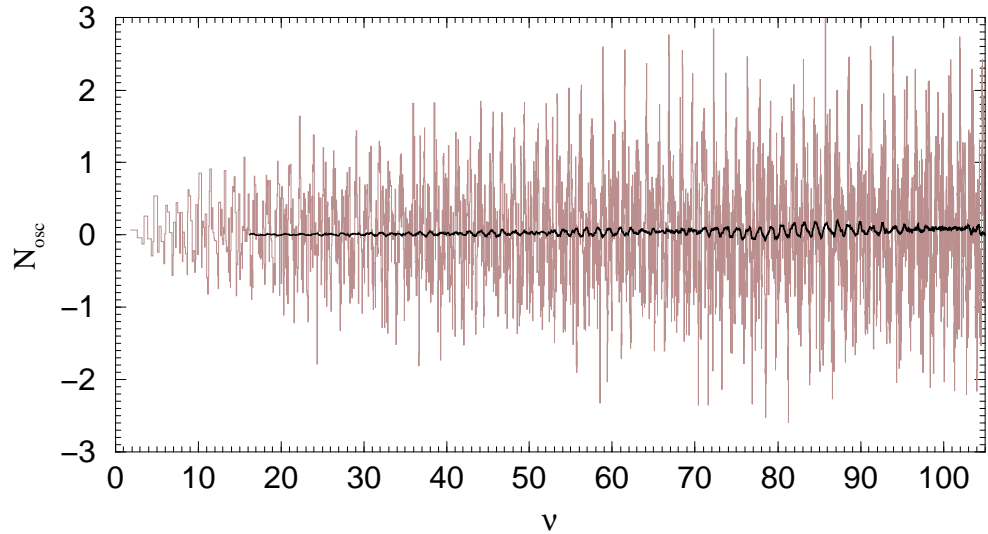} 
\caption{%
Fluctuating part of the spectral staircase, ${\rm N}_{\rm
fluc}\defas{\rm N}(\nu)-\overline{\rm N}$ for the asymmetric stadium at
$\rho=1.2$.  The displayed range contains the first $5000$ points in
the spectrum.  The heavy line is a running average over 250
neighboring points.  
}
\label{fig:Nosc}
\end{figure}

In the following we demonstrate the performance of the described
method by exhibiting some numerical results on magnetic billiards
which have been inaccessible by other methods.

\subsection{Spectral statistics}
We start by considering spectral statistics based on large data sets
of calculated spectral points.  As explained in the introduction, the
spectra are defined in the semiclassical direction $b\to0$ keeping the
cyclotron radius $\rho$ constant. This way the underlying classical
dynamics is fixed. For classically hyperbolic systems one expects
Random Matrix Theory (RMT) to reproduce the spectral statistics.

We use the two domains described in Fig.~\ref{fig:shapes}. One is an
asymmetric version of the Bunimochwich stadium billiard ($r_1=0.75,
r_2=0.25, \Area=5.39724$). In the magnetic field its dynamics is free
of unitary symmetries but contains an anti-unitary one (time reversal
and reflection at $y=0$.) On the other hand, the skittle shape (made
up of the arcs of four symmetrically touching circles, $r_1=1.0,
r_2=0.5, \Area=4.33969$) does not have any symmetry. We choose it
because it generates hyperbolic classical motion even for small
cyclotron radii $\rho>1$, according to a recent theorem
\cite{Gutkin99}. The asymmetric stadium could not be proven to be
hyperbolic although we find no numerical evidence for systematic
deviations from the RMT behavior (see below.)

We calculated 5300 and 7300 consecutive interior Dirichlet eigenvalues
at $\rho=1.2$ for the asymmetric stadium and the skittle shaped
domain, respectively.  It should be noted, that states of much
higher ordinal number can be computed at little cost with the
present method. The time consuming task is rather to find
\emph{all} energies $\nu_i=\rho^2/b_i^2$, including the near-degenerate
ones, in a given interval.

A quantity which sensitively indicates whether spectral points were
missed is the fluctuating part, ${\rm N_{fluct}}$, of the spectral
staircase function
\begin{eqnarray}
 {\rm N}(\nu) \defas \sum\nolimits_i \Theta(\nu-\nu_i) 
 = \overline{\rm N}(\nu)+{\rm N_{fluct}}(\nu).
\end{eqnarray}
As shown recently \cite{MMP97}, its smooth part coincides with the
non-magnetic one in its leading terms. In our units and for Dirichlet
boundary conditions they read
\begin{eqnarray}
 \overline{\rm N}(\nu) = \frac{\Area}{\rho^2\pi} \nu^2
	- \frac{\Len}{2\pi\rho} \nu + \frac{1}{6},
\end{eqnarray}
where $\Area$ is the domain area and $\Len$ the boundary length. The
constant, which contains an integral over the boundary curvature, is
the same for the shapes considered.  Figure \ref{fig:Nosc} displays
the fluctuating part of the number function ${\rm N}_{\rm
fluc}\defas{\rm N}(\nu)-\overline{\rm N}$ for the asymmetric stadium.
It proves that the spectrum is complete.  A similar result is
obtained for the skittle shaped domain (not shown.)

\begin{figure}
\epsfxsize\textwidth%
\epsfbox{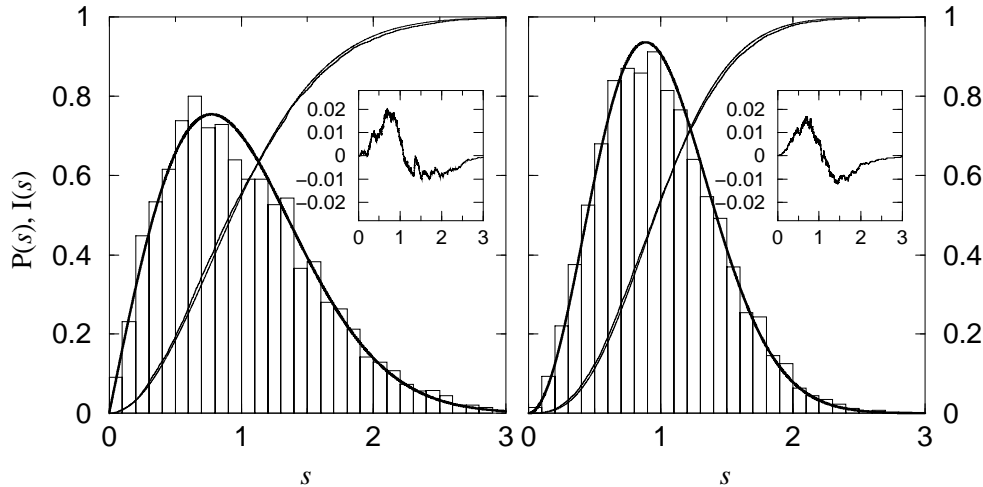} 
\caption{%
Nearest neighbor distributions of the asymmetric stadium (left) and
the skittle shaped domain (right), at $\rho=1.2$.  The histograms
should be compared to GOE and GUE predictions of Random Matrix Theory,
respectively (heavy lines.)  The monotonic lines give the
corresponding cumulative quantities. 
Their differences are reported in the insets.
}
\label{fig:NN}
\end{figure}

The large spectral intervals at hand allow us to calculate directly
some of the popular spectral functions.  Due to the underlying
classical chaos and the symmetry properties mentioned above one
expects the statistics of the Gau{\ss}ian Orthogonal Ensemble (GOE)
for the asymmetric stadium and the Gau{\ss}ian Unitary Ensemble (GUE)
for the skittle.  Figure \ref{fig:NN} shows the distributions of
nearest neighbors $P(s)$ of the unfolded spectra. Indeed, one finds
excellent agreement with Random Matrix Theory. The differences between
the numerical and the RMT cumulative functions $I(s)=\int_0^s
P(s^\prime) \rmd s^\prime$ stay below $2\%$.

\begin{figure}
\epsfxsize\textwidth%
\epsfbox{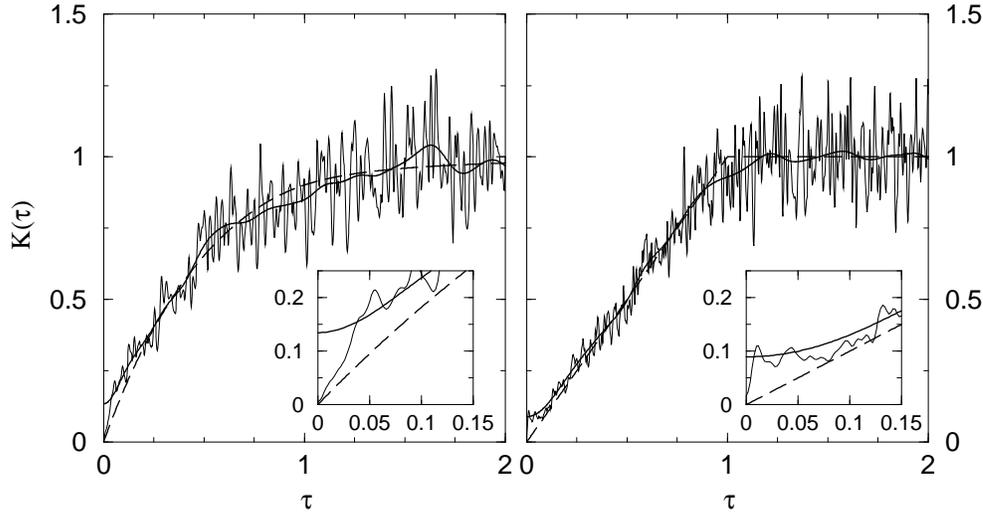} 
\caption{%
Spectral form factor of the asymmetric stadium (left) and the skittle
shaped domain (right), based on 5300 and 7300 spectral points,
respectively.  The heavy lines display the same data after stronger
spectral averaging.  The random matrix result for the Gau{\ss}ian
Orthogonal and the Gau{\ss}ian Unitary Ensemble, respectively, is
indicated by the dashed lines.
The insets show the regions of small $\tau$.
}
\label{fig:formfactor}
\end{figure}

A function which characterizes the 
spectrum much more sensitively than $P(s)$ is the form factor $
K(\tau)$, i.e. the (spectrally averaged) Fourier transform of the
two-point autocorrelation function of the spectral density
\cite{Berry85,AIS92}.
Figure \ref{fig:formfactor} gives the spectral form factor together
with the RMT results. We find very good agreement. One would expect
systematic deviations at small $\tau$ due to the contributions of
single short periodic orbits. These cannot be resolved with the
present size of the spectral interval, though.  Since most other
popular spectral measures like Dyson's $\Delta_3$ statistic are
functions of the form factor there is no need to present them here.

The good agreement with RMT is not only a consequence of the
large spectral intervals the statistics are based on. It is equally
important that the spectra are defined at fixed classical dynamics.
Had we calculated the spectra at fixed field, they
would have been based on a classical phase space that transforms from
a near-integrable, time-invariance-broken structure to a hyperbolic
time-invariant one as $\rho$ increases with energy.  This
transformation of spectral statistics from GOE to GUE as the field is
increased was studied in \cite{BGOdAS95,YH95,ZLB95}.

%
%
%
%
%

%
%
\begin{figure} 
\centerline{%
\epsfxsize0.46\textwidth%
\epsfbox{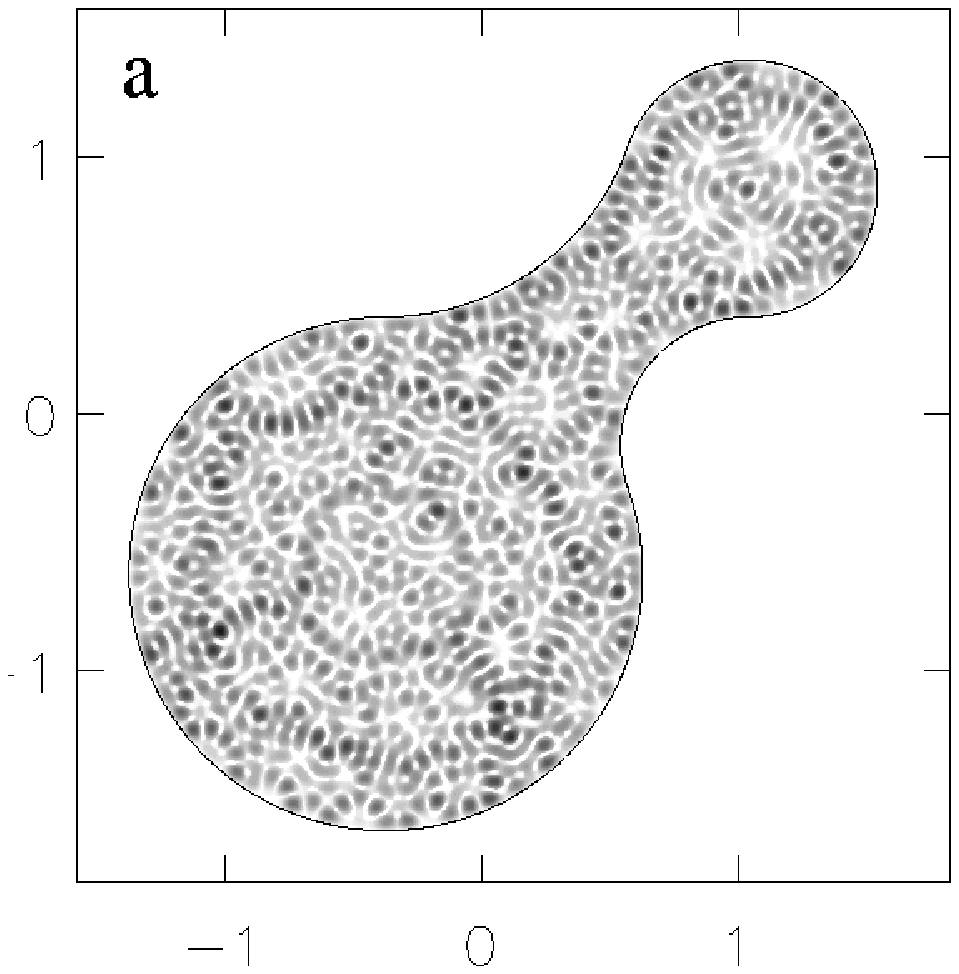} 
\hfill
\epsfxsize0.46\textwidth%
\epsfbox{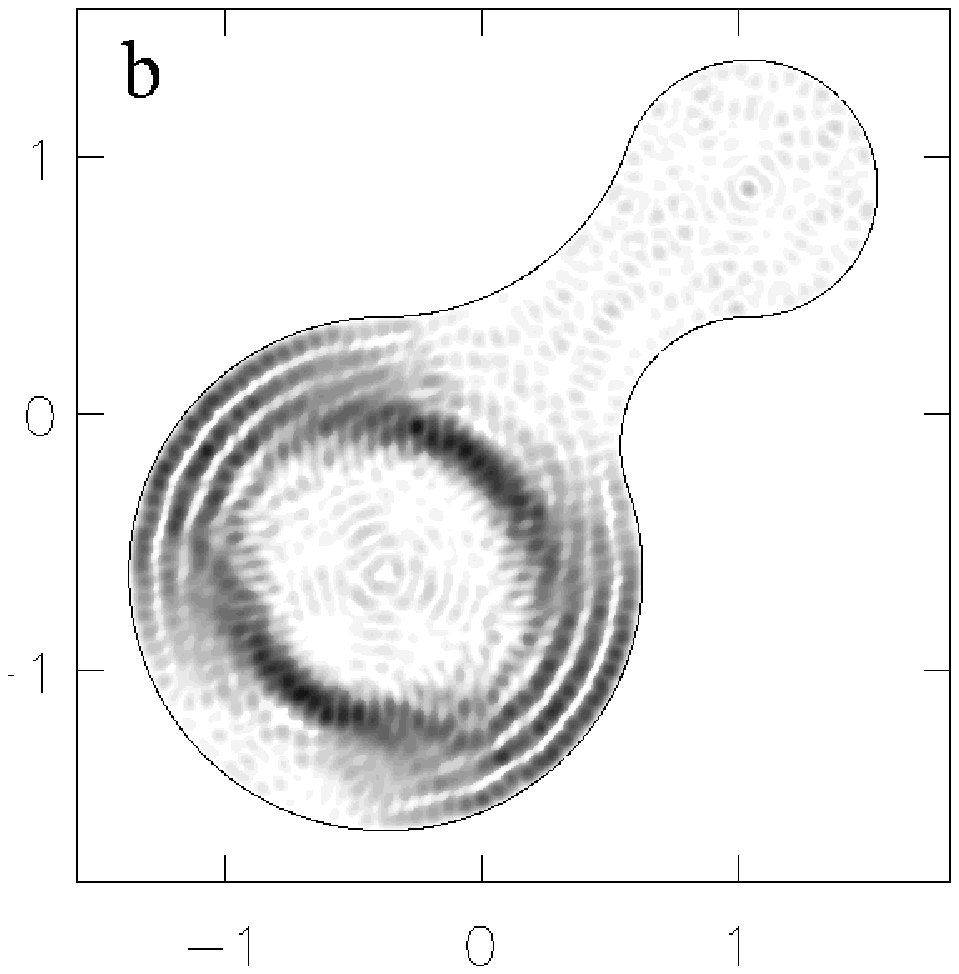} 
}
\vspace*{2ex}
\centerline{%
\epsfxsize0.46\textwidth%
\epsfbox{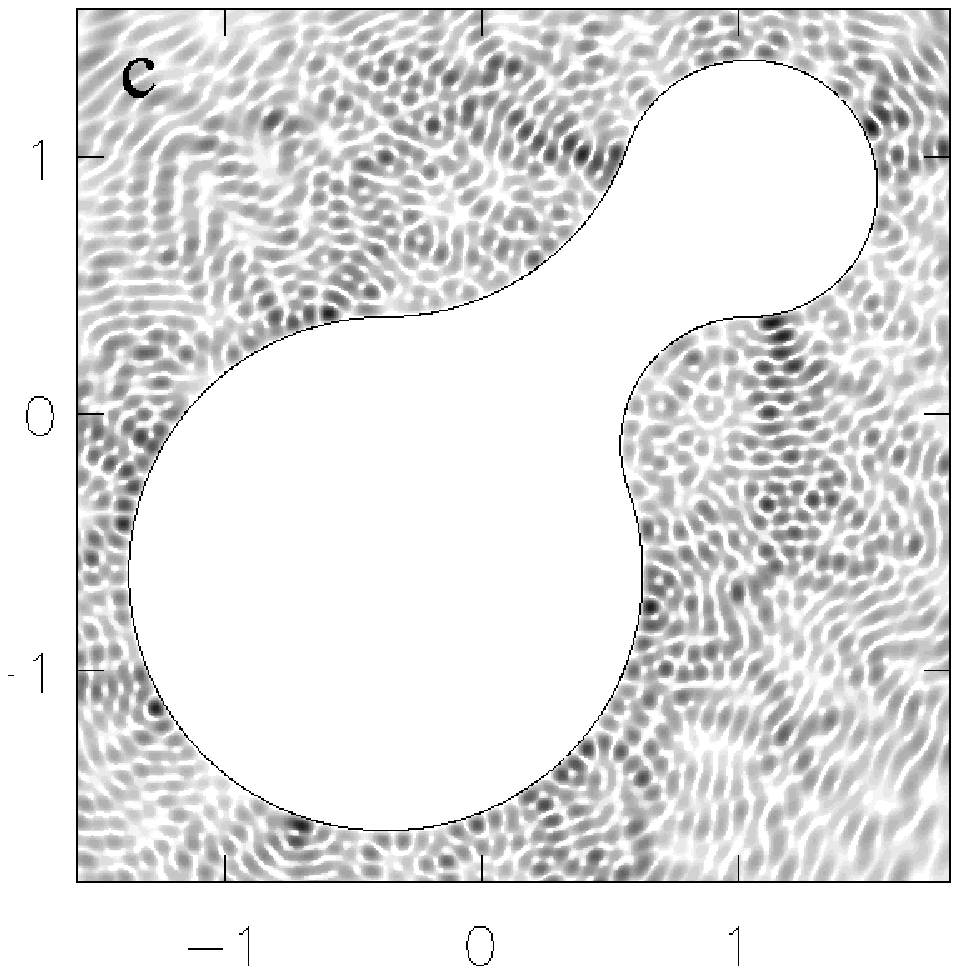} 
\hfill
\epsfxsize0.46\textwidth%
\epsfbox{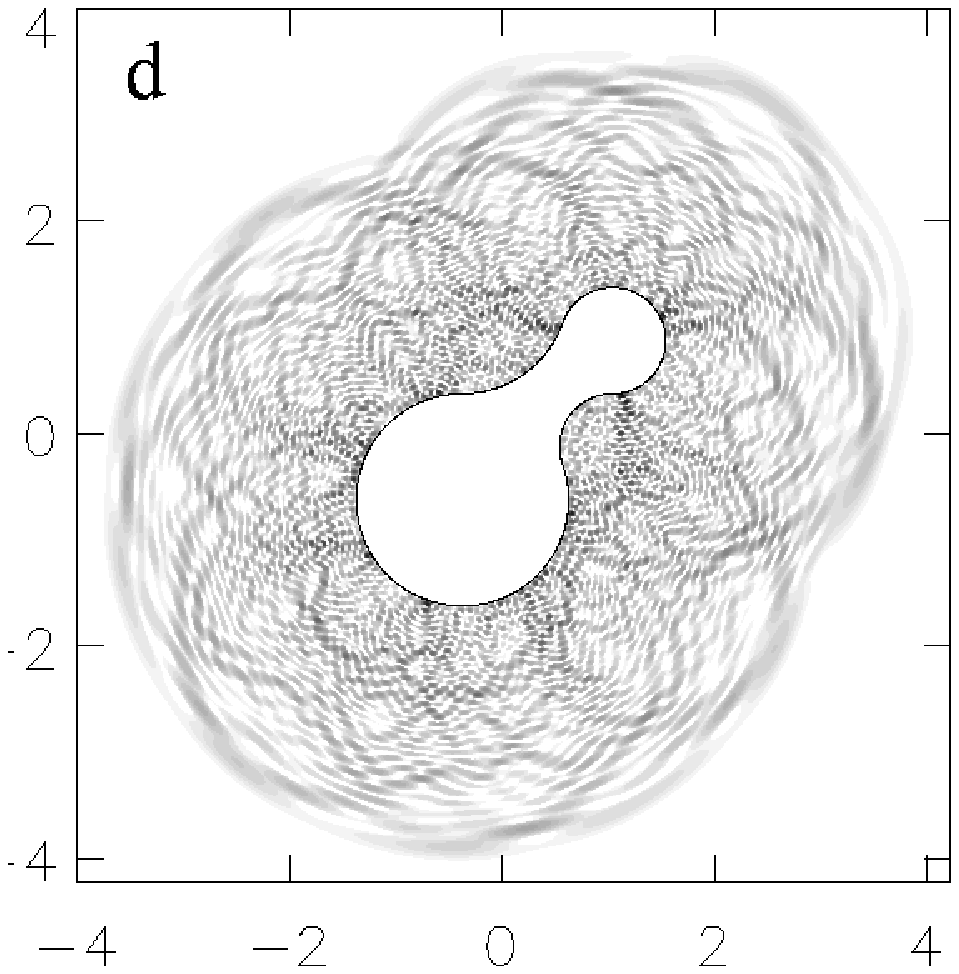} 
}
\vspace*{2ex}
\centerline{%
\epsfxsize0.46\textwidth%
\epsfbox{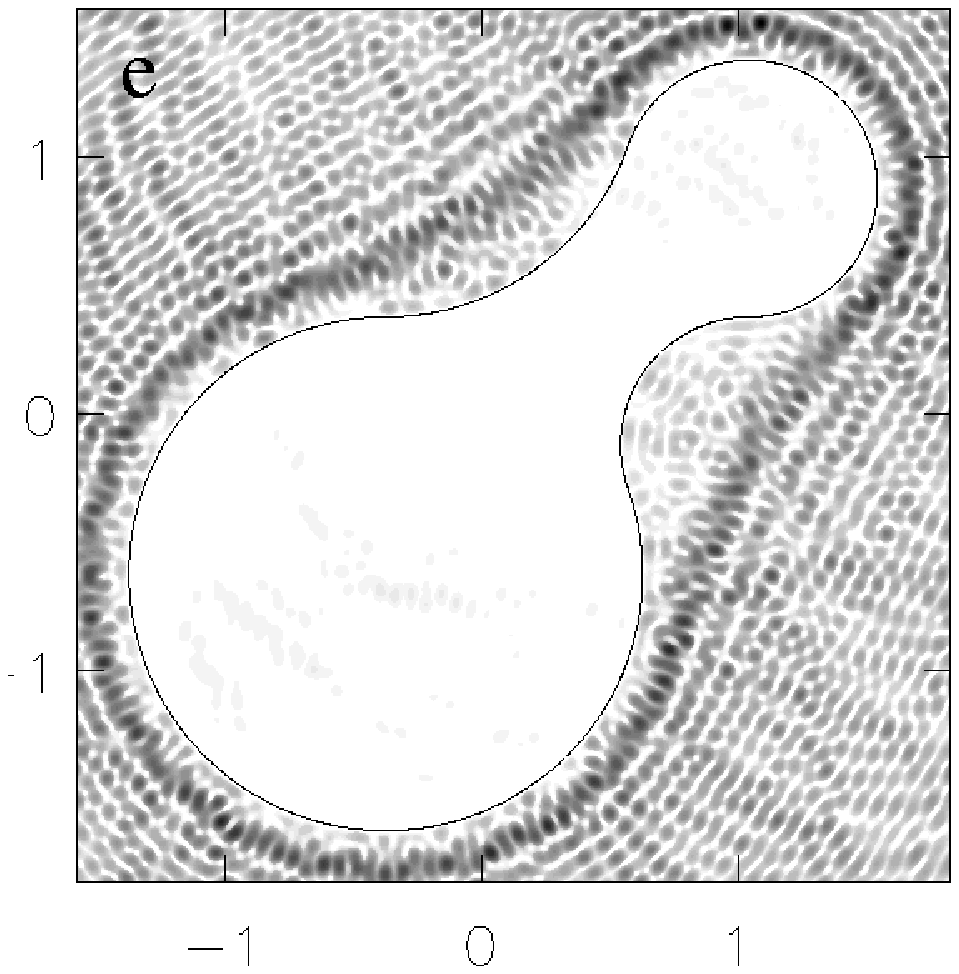} 
\hfill
\epsfxsize0.46\textwidth%
\epsfbox{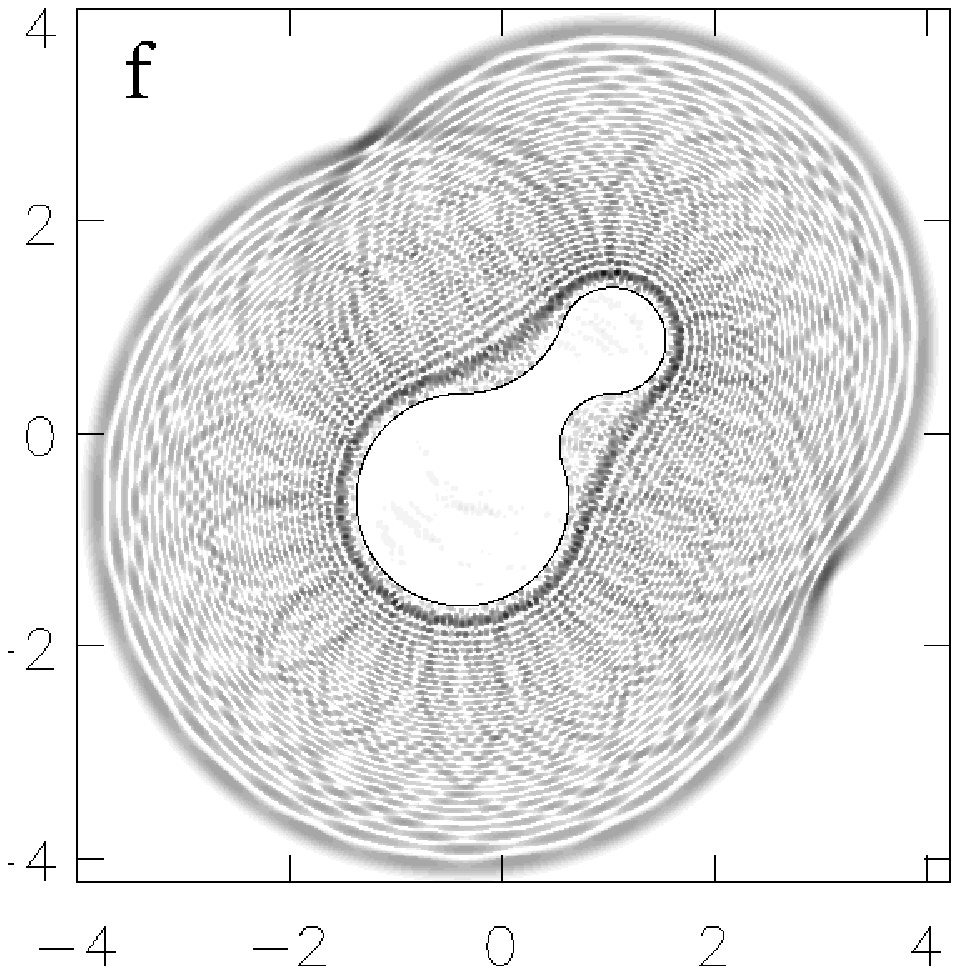} 
}
\caption{%
Interior and exterior wave functions of the skittle shape at
$\rho=1.2$ around the one-thousandth interior eigenstate. The plotted
shade is proportional to $|\psi|$, the thin lines indicate the
boundary $\Gamma$. Energies: (a) $\nu\simeq32.98804$, (b)
$\nu\simeq33.12033$, (c,d) $\nu\simeq32.84740$, (e,f)
$\nu\simeq32.50073$.}
\label{fig:skittle}
\end{figure}

%
%
\begin{figure} 
\centerline{%
\epsfxsize0.60\textwidth%
\epsfbox{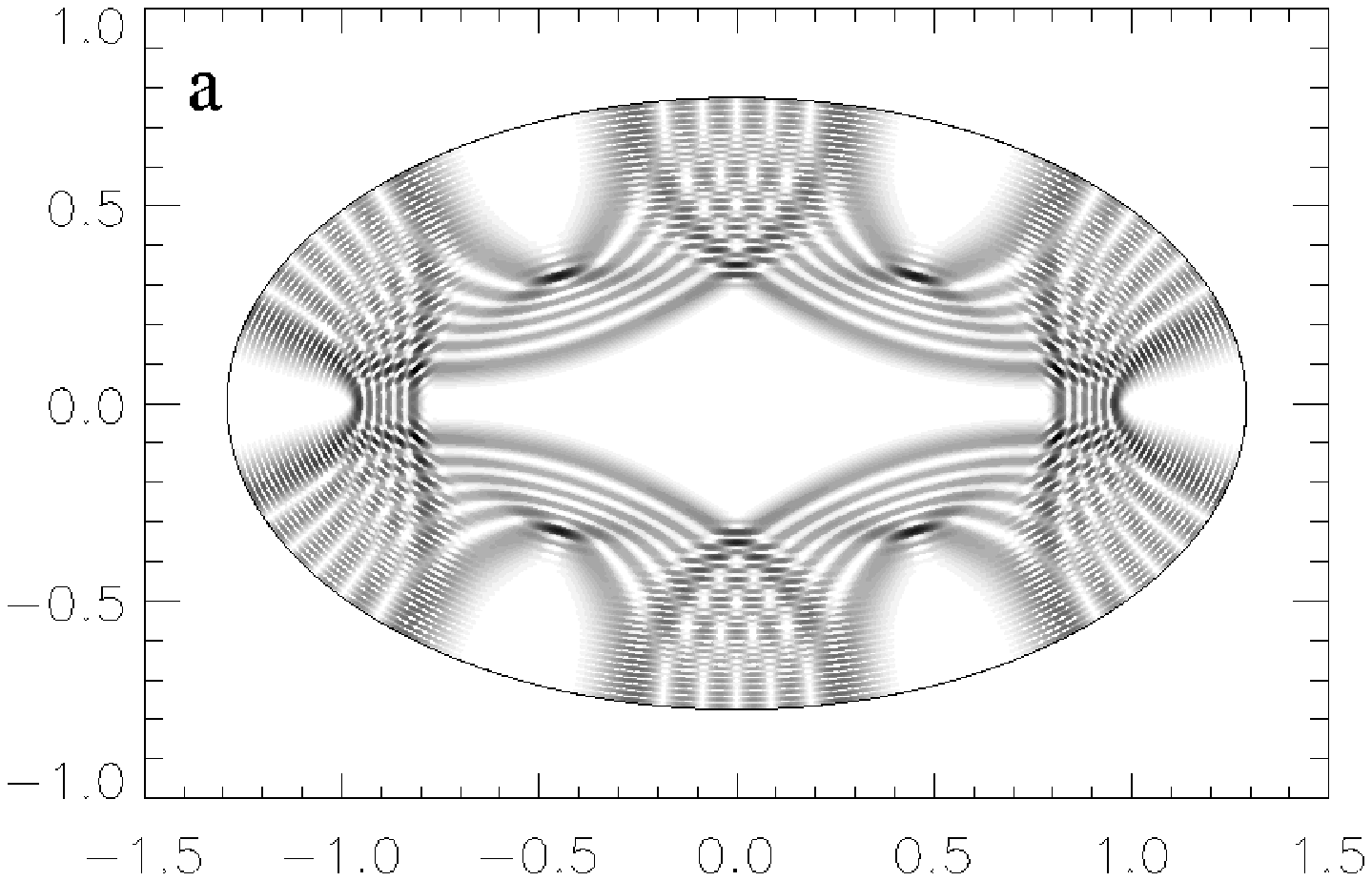} 
}
\vspace*{2ex}
\centerline{%
\epsfxsize0.60\textwidth%
\epsfbox{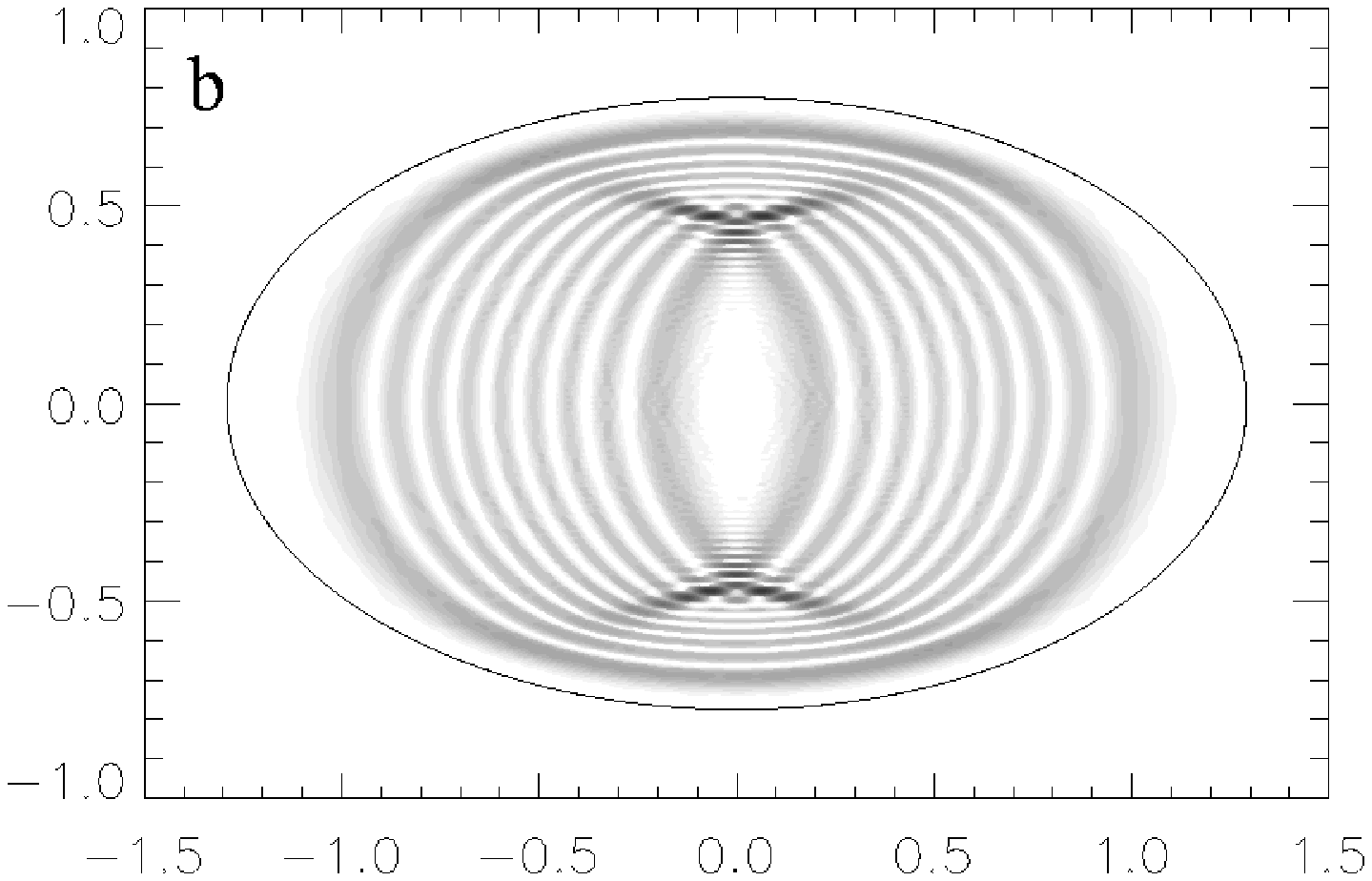}
}
\vspace*{2ex}
\centerline{
\epsfxsize0.5\textwidth%
\epsfbox{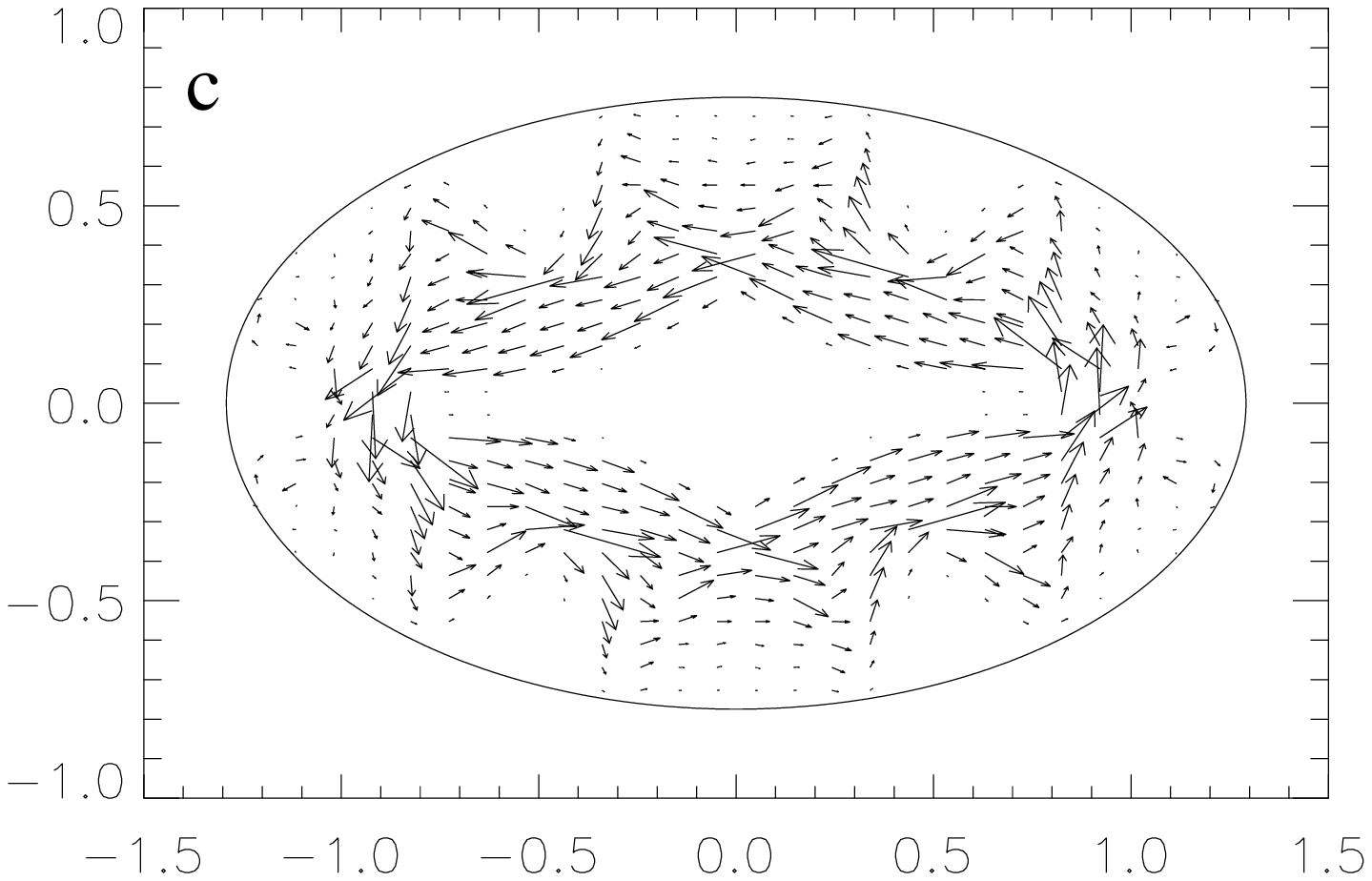} 
\hfill
\epsfxsize0.5\textwidth%
\epsfbox{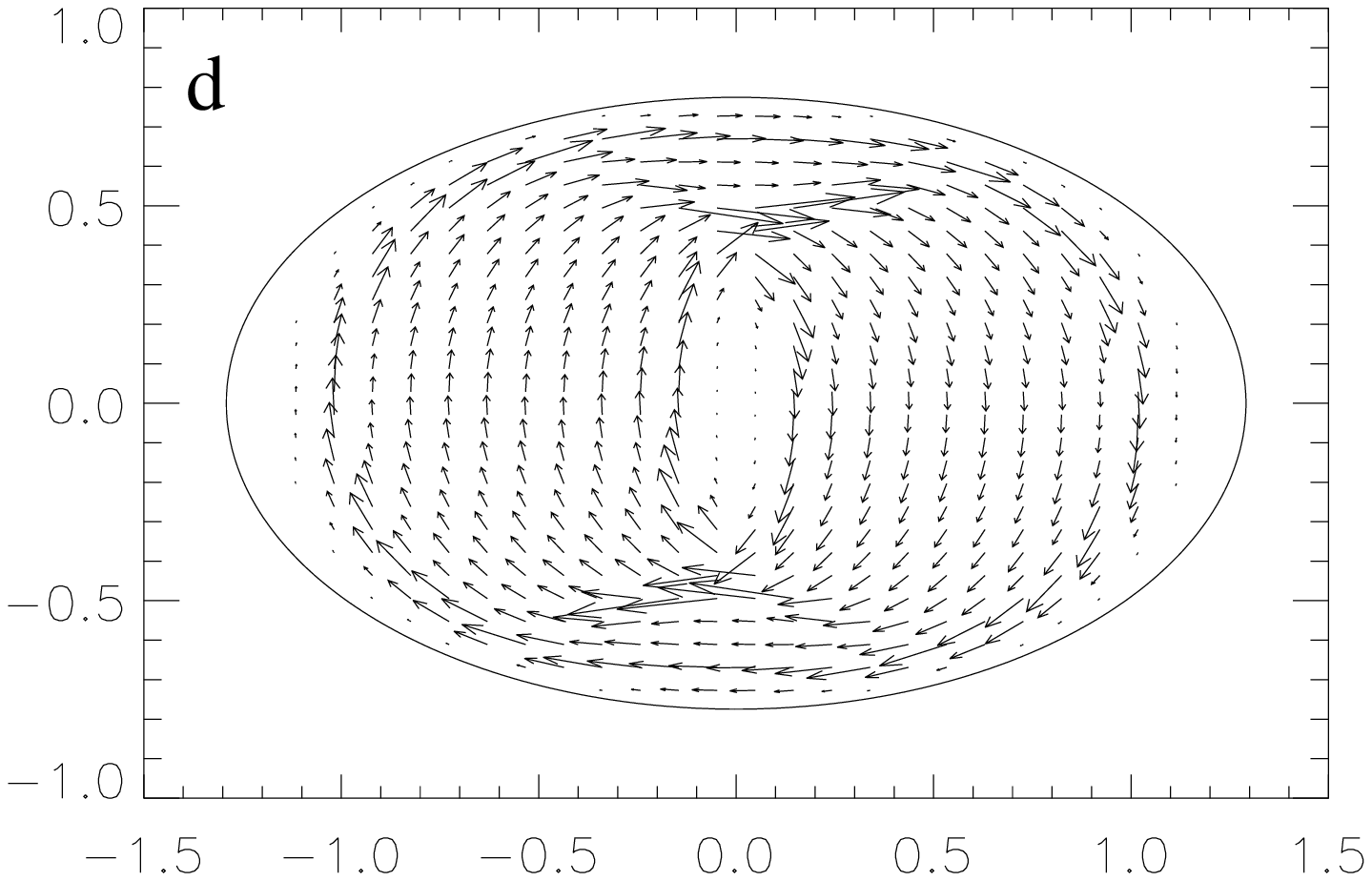} 
}
\caption{%
Wave functions (a,b) and current distributions (c,d) in an elliptic
domain at $\rho=0.6$, around the ten-thousandth interior eigenstate,
with energies $\nu\simeq60.06026$ (a,c) and $\nu\simeq60.50030$
(b,d).  }
\label{fig:ellipseint}
\end{figure}

%
%
\begin{figure}
\centerline{%
\epsfxsize0.60\textwidth%
\epsfbox{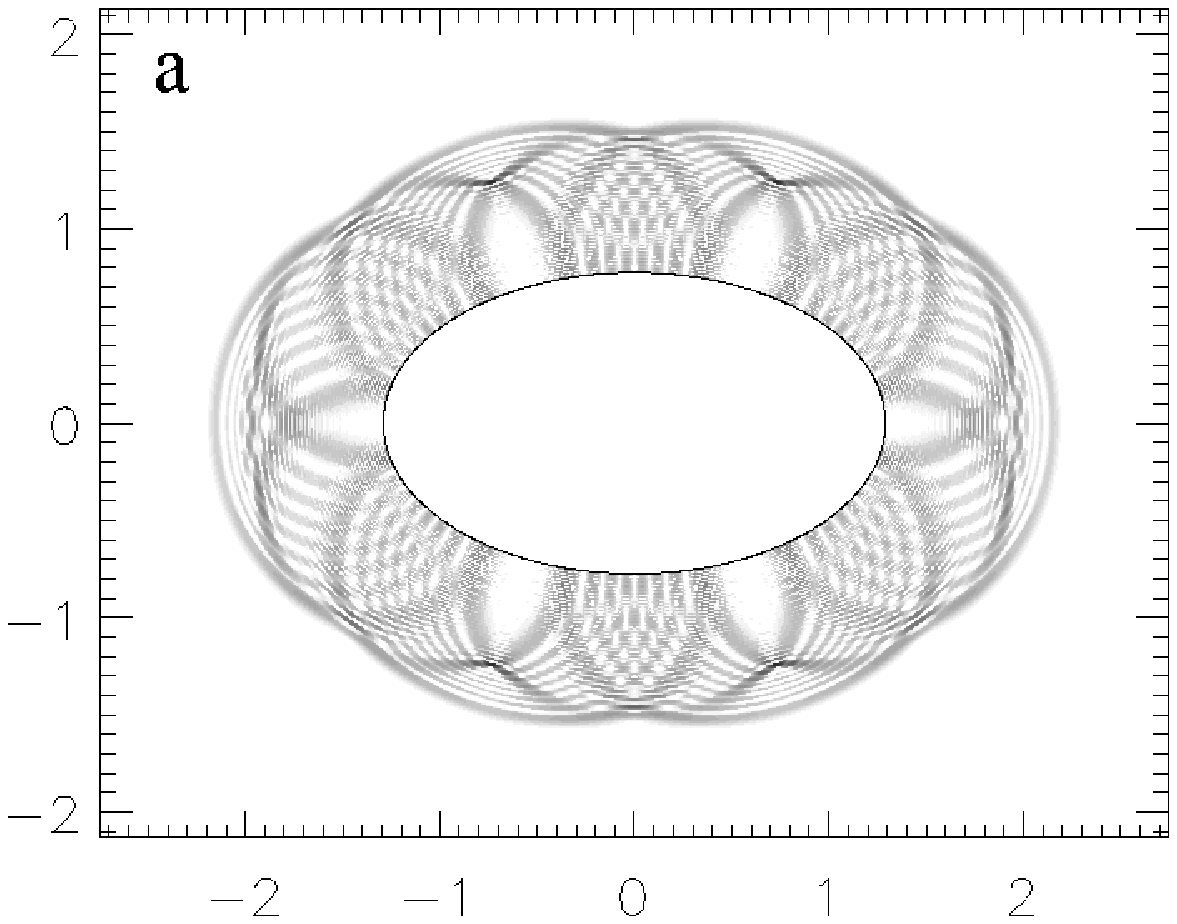} 
}
\vspace*{2ex}
\centerline{%
\epsfxsize0.60\textwidth%
\epsfbox{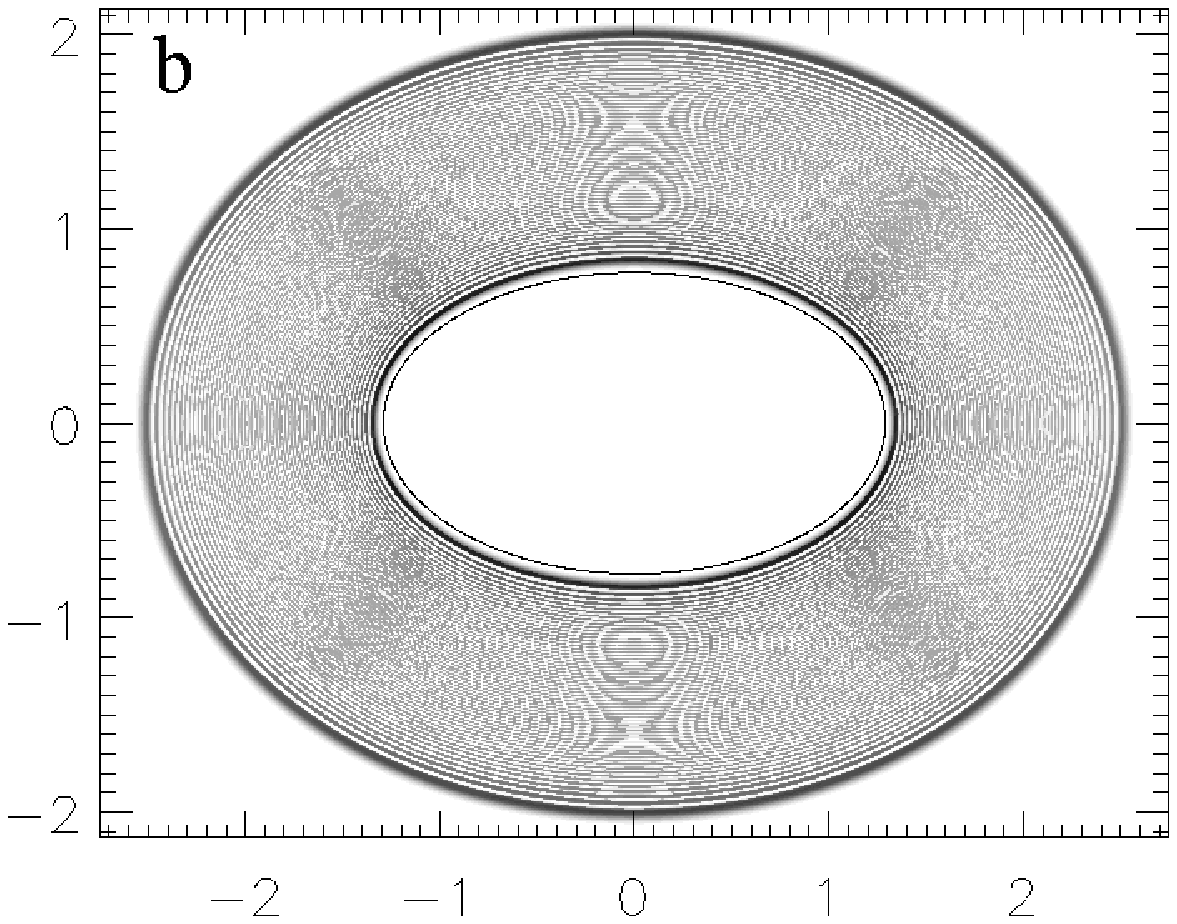}
}
\vspace*{2ex}
\centerline{
\epsfxsize0.5\textwidth%
\epsfbox{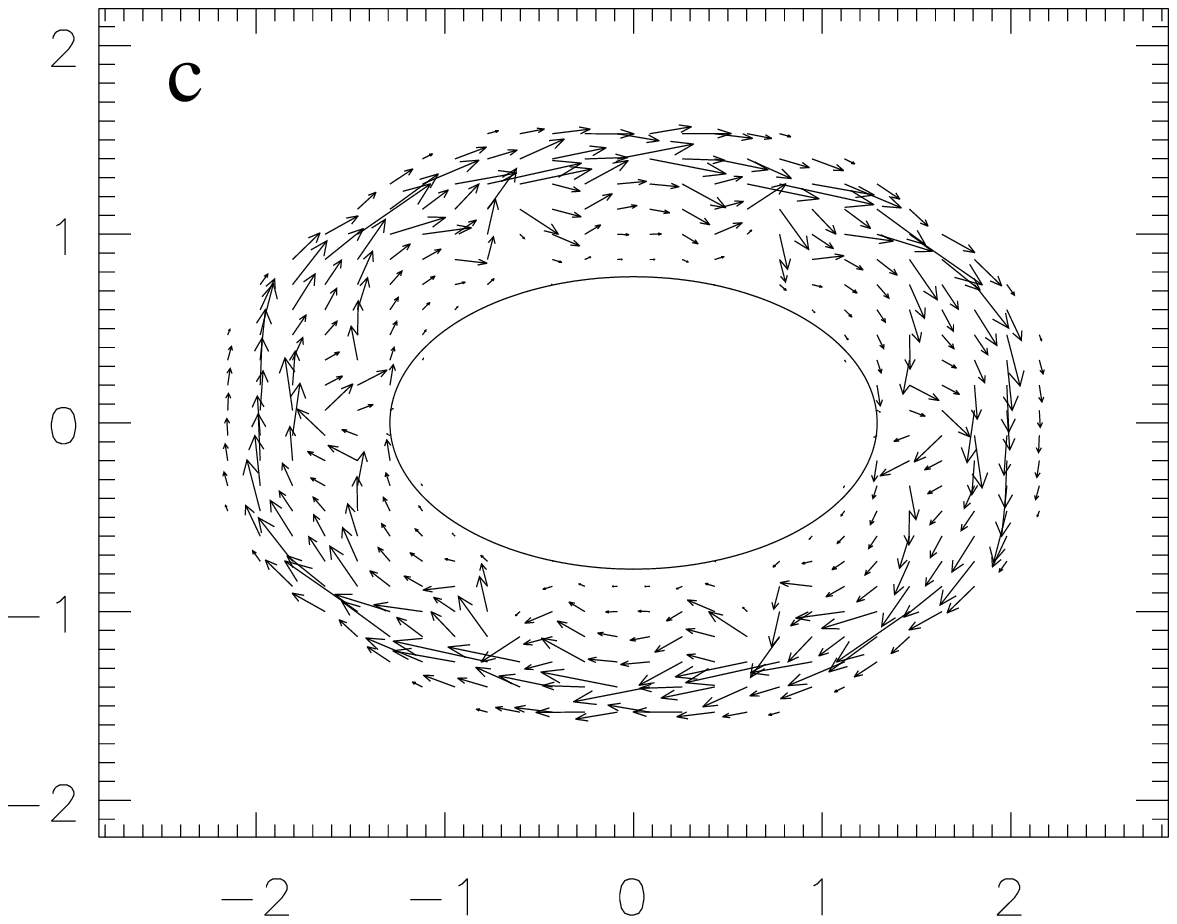} 
\hfill
\epsfxsize0.5\textwidth%
\epsfbox{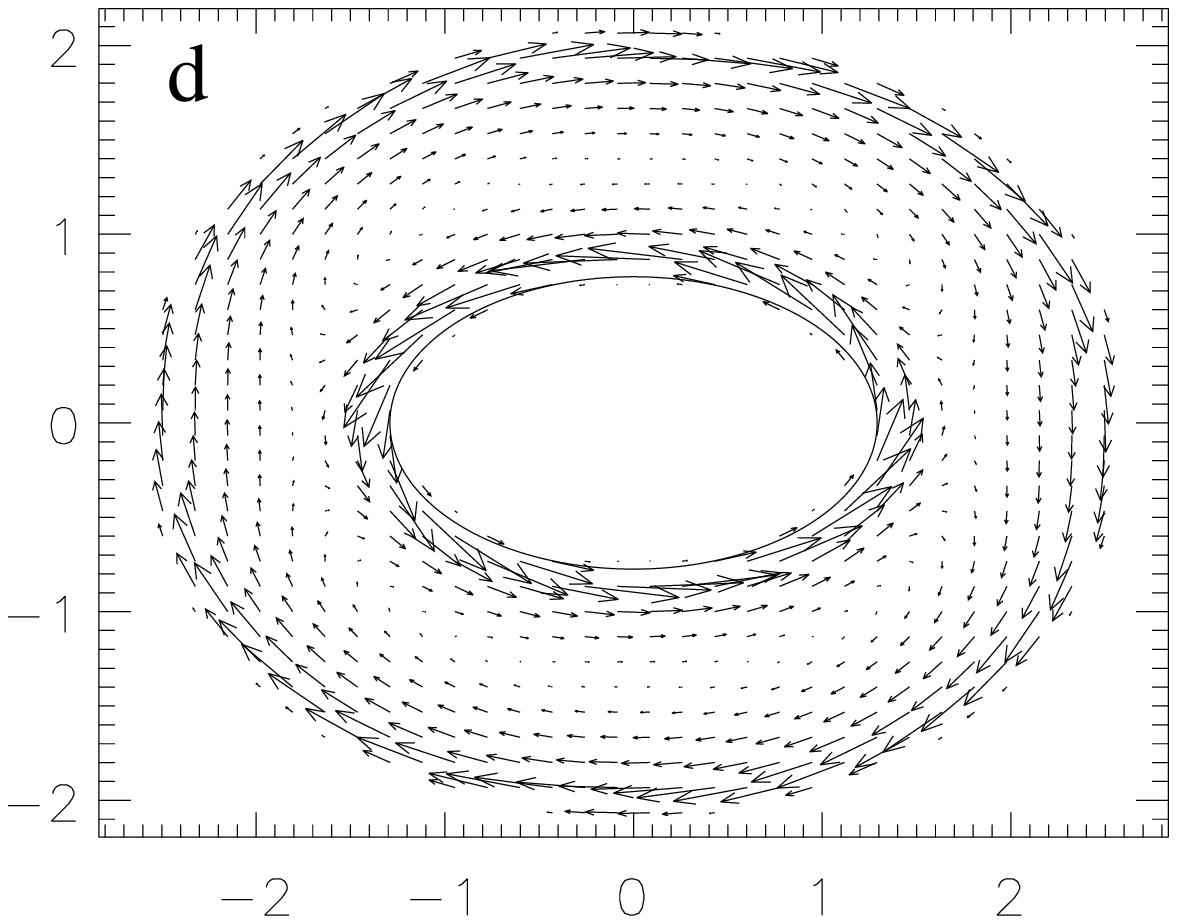} 
}
\caption{%
Exterior wave functions (a,b) and current distributions (c,d) at
$\rho=0.6$ and at similar energies as in Fig. \ref{fig:ellipseint},
$\nu\simeq60.13634$ (a,c) and $\nu\simeq60.50049$ (b,d).  }
\label{fig:ellipseext}
\end{figure}

\subsection{Wave functions}

We proceed to present a selection of wave functions calculated in the
semiclassical regime. We start with those of the skittle shaped domain
choosing again $\rho=1.2$. This ensures that the corresponding
classical skipping motion is hyperbolic in the interior, as well as in
the exterior.  

Figure \ref{fig:skittle}(a) shows the density plot of a typical
interior wave function around the one-thousandth eigenstate.
As expected for a classically ergodic system, it spreads out
throughout the whole domain but is not completely featureless.
Occasionally, one may also find \emph{bouncing-ball} modes, i.e. wave
functions localized on a manifold of marginally stable periodic
orbits. One such wave function is given in
Fig. \ref{fig:skittle}(b). It belongs to a family of 2-orbits.

A typical \emph{exterior} wave function with an energy close to that
of Fig. \ref{fig:skittle}(a) is displayed in the middle row of Figure
\ref{fig:skittle}, at the same scale (c) and a larger scale (d).  One
observes that in the vicinity of the boundary it behaves quite similar
to an interior function.  On a larger scale, the wave function decays 
after a distance smaller than two cyclotron radii. In this region
circular structures are faintly visible with the radius of the
classical cyclotron motion.

The bottom row of Figure~\ref{fig:skittle} shows a quite different
exterior state with an energy close to that of a Landau level.  It is
a bulk state. A typical feature is the fact there are no large
amplitudes close to the boundary. Rather, one finds a ring of
increased amplitude encircling the domain. Another ring surrounds the
domain at a distance of $ 2 \rho$. This double-ring structure moves
outwards as one goes through the series of states with energies
increasingly close to the Landau levels.  Semiclassically, it can be
understood as being made up of a superposition of cyclotron
orbits. This becomes even more clear in the following where we
consider a more symmetric shape of the boundary.

For the second set of wave functions we choose an elliptic domain (of
eccentricity $0.8$ and area $\pi$) at a small cyclotron radius
$\rho=0.6$.  The classical dynamics is mixed chaotic in this case
\cite{RB85}. Going to the extreme semiclassical limit -- the
ten-thousandth interior eigenstate -- we expect the wave functions to
mimic the structures of the underlying classical phase space.

Indeed, Figure \ref{fig:ellipseint}(a) displays a wave function which
is   localized along a stable interior $6\times6$-orbit. Note
that the wave nature of the eigenstate is still visible at points
where the trajectory crosses with itself, in particular at the shallow
intersections close to the center.

Since $\rho$ is small enough to allow closed cyclotron orbits within
the ellipse, we find bulk states also in the interior, see Fig.
\ref{fig:ellipseint}(b) for an example.  Again it is semiclassically
described by a superposition of closed cyclotron orbits. This can be
seen clearly from the current distributions which are given in the
bottom row of Fig. \ref{fig:ellipseint} for the edge state (c) and the
bulk state (d), respectively. Here, the length of the arrows is
proportional to the amplitude of the current density.

Similar semiclassical states can also be found in the exterior, as
displayed in Figure \ref{fig:ellipseext}.  The edge state,
Fig. \ref{fig:ellipseext}(a), obviously belongs to an
$8\times6$-orbit. Like all edge states it is distinguished from a
typical bulk state, cf. Fig. \ref{fig:ellipseext}(b), by the finite
current it carries around the domain. In contrast, the bulk state with
its counter-running current densities has no net current along the
boundary, cf. Fig.~\ref{fig:ellipseext}(c) and
Fig.~\ref{fig:ellipseext}(d).

We emphasize that all the wave functions and current distributions
shown above are calculated throughout the entire displayed area.  They
turn out to be \emph{numerically} zero in the complementary domains as
expected from the theory.  Consequently, the type of a solution of a
\emph{single} integral equation can be inferred by calculating the
wave function.

%
%
\begin{figure}
\centerline{%
\epsfxsize0.7\textwidth%
\epsfbox{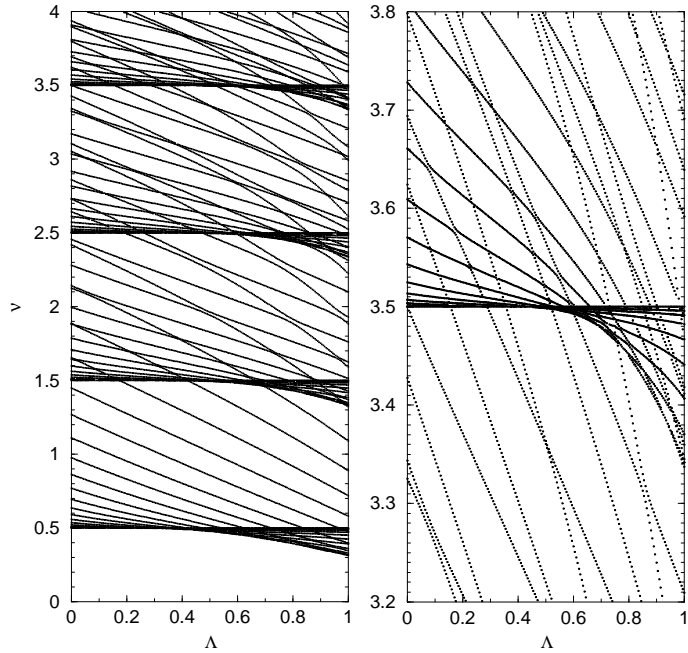}
}
\caption{%
Parametric dependence of the \emph{exterior} spectrum on the boundary
condition (for the asymmetric stadium at fixed $b=0.25$).  The
parameter $\Lambda$ interpolates between Dirichlet ($\Lambda=0$) and
Neumann ($\Lambda=1$) boundary conditions.  The right graph shows
details around the forth Landau level.}
\label{fig:lambdadyn}
\end{figure}

\subsection{General boundary conditions}

So far, only Dirichlet boundary conditions were considered. As a last
point we show the parametric dependence of a spectrum on the type of boundary
conditions. Figure \ref{fig:lambdadyn} presents the \emph{exterior}
spectrum of the asymmetric stadium, calculated at fixed $b=0.25$.
The value $\lambda$ is taken constant along the boundary and parameterized by
a number $\Lambda\in [0;1]$,
\begin{eqnarray}
 \lambda=-\frac{\rho}{2\nu}\tan\big(\tfrac{\pi}{2}\Lambda\big).
\end{eqnarray}
Here $\lambda$ is chosen negative to ensure that the transformation
from Dirichlet ($\Lambda=0$) to Neumann ($\Lambda=1$) boundary
conditions is continuous. For positive $\lambda$ this would not be the
case, which is a restriction similar to the one for the field free
case \cite{SPSUS95}.

The energies clustering around the Landau levels
$\nu_n=n+\tfrac{1}{2}, n\in\mathbb{N}_0$ belong to bulk states.  One
observes that they are lifted from the Landau levels into higher
energies at Dirichlet boundary conditions, whereas in the Neumann case
they are shifted to smaller energies.  A semiclassical theory which
describes the exponential approach of the bulk states to the Landau
levels and its transition as a function of $\Lambda$ will be published
elsewhere.

%
%
%
\section*{Conclusions}

The main theoretical result presented here is the finding that the
two boundary integral equations of the billiard problem admit spurious
solutions in the magnetic case, and how those are identified and
removed.  

An important implication concerns the semiclassical theory of magnetic
billiards.  The trace formula for the semiclassical quantization of
the field free case is based on a (double layer) boundary integral
equation \cite{BB72}.  If one tries to repeat the derivation for the
magnetic case, problems should arise since the equation does not give
a sufficient condition.  Indeed, a (Balian-Bloch-like) derivation of
the trace formula for magnetic billiards does not exist.  We emphasize
that the starting point should be the \emph{gauge-invariant}
formulation of the boundary integral equation, as presented above.
Only then are the spurious solutions gauge invariant, have a physical
interpretation, and can be taken into account systematically.

We have shown how a precise and efficient computational
method for the calculation of spectra and wave functions can be based
on a combination of boundary integral equations. It allows to obtain
the exact spectra and wave functions at energies and fields
inaccessible hitherto.

The possibility of calculating the exterior spectra as well, brings up
the question on how interior and exterior spectra are related.  Here,
a problem is the existence of an infinity of bulk states which do not
have much physical relevance but prevent the spectral number function
from being well-defined.  Our calculation of the exterior level
dynamics as a function of the boundary condition shows that the edge
states and bulk states differ in their sensitivity to the boundary
condition.  In a forthcoming communication we will propose a
definition of the spectral edge state density based on this
observation.

\section*{Acknowledgments}

It is a pleasure to thank B Gutkin and M Sieber for fruitful
discussions. The work was supported by a Minerva fellowship for KH,
and by the Minerva Center for Nonlinear Physics.

%
%
%
\setcounter{section}{0}
\setcounter{equation}{0}
\renewcommand{\theequation}{A.\arabic{equation}}
%
%
%
\section*{Appendix A. The free magnetic Green function}

The free magnetic Green function was derived in \cite{KR92,TCA97} by
angular momentum decomposition.  Here, we show how it is obtained by
directly performing the Fourier transform of the time evolution
operator \cite{FH65}
\begin{eqnarray}
\fl
  {\rm U}(\rvec,\rvec_0;t)
  =
  \frac{1}{2\pi\rmi}\frac{1}{b^2} \frac{1}{\sin(\omega t)}\,
  \rme^{\ts\rmi\big
            (\tfrac{(\rvec-\rvec_0)^2}{2 b^2}\cot(\omega t)
             -\tfrac{\rvec\times\rvec_0}{b^2}+\chit(\rvec)-\chit(\mb{r_0})
            \big)}
\end{eqnarray}
which yields both, the gauge dependent and the gauge independent part
in a straightforward manner.  We have to evaluate
\begin{eqnarray}
\fl
  {\rm G}(\rvec;\rvec_0)=
  \frac{b^2}{2\rmi}\int_0^\infty  
  \!\!\!\!  {\rm d}(\omega t)\,\,
  \rme^{\ts\rmi E t/\hbar}\,
  {\rm U}(\rvec,\rvec_0;t)\,
\nnn
=
  \rme^{\ts -\rmi(\tfrac{\rvec\times\rvec_0}{b^2}-\chit(\rvec)+\chit(\mb{r_0}))}
 \,{\rm G}^0_\nu\big(\tfrac{(\rvec-\rvec_0)^2}{b^2}\big)
\end{eqnarray}
assuming that the energy $\nu$ has a small imaginary part to ensure
convergence. For the gauge independent part one obtains
\begin{eqnarray}
\label{eq:contint}
\fl
 {\rm G}^0_\nu(z) = \frac{-1}{4\pi} 
 \int_0^\infty\!\! \frac{{\rm d}\tau}{\sin(\tau)}\,
  \rme^{\ts\rmi(z\cot(\tau)/2+2\nu\tau)} 
 \nnn
\fl
= 
 \frac{-1}{4\pi} \sum_{n=0}^\infty  \rme^{\ts2\pi\rmi\nu n}
 \int_0^\pi\!\! 
 \frac{{\rm d}\tau}{\sin(n\pi+\tau)}\,
  \rme^{\ts\rmi(z\cot(n\pi+\tau)/2+2\nu\tau)} 
 \nnn
\fl
= 
 \frac{-1}{4\pi}
 \frac{1}{1+\rme^{2\pi\rmi\nu}} 
   \bigg\{
     \int_{0}^\infty\!\!\!\!
     \frac{{\rm d}u}{\sqrt{1+u^2}} \Big(\frac{u+\rmi}{u-\rmi}\Big)^\nu
     \rme^{\rmi z u/2} 
    +\rme^{2\pi\rmi\nu}
     \int_{-\infty}^0
     \frac{{\rm d}u}{\sqrt{1+u^2}} \Big(\frac{u+\rmi}{u-\rmi}\Big)^\nu
     \rme^{\rmi z u/2} 
   \bigg\}
 \nnn
 \fl
= 
 \frac{-1}{4\pi}\,
 \Gamma(\tfrac{1}{2}-\nu) \,
  \bigg[
 \rme^{-\rmi\pi(\nu-\frac{1}{2})}\, \frac{\Gamma(\tfrac{1}{2}+\nu)}{2\pi\rmi}
 \bigg\{
  \int_{0}^{-\rmi\infty}\!\!\!\!\!\!\!\!\!\!\!{\rm d}t\,
    (t+1)^{\nu-\frac{1}{2}} (t-1)^{-\nu-\frac{1}{2}} \rme^{- z t/2} 
  \nnn
  +\rme^{2\pi\rmi\nu}\!\!
  \int_{+\rmi\infty}^0\!\!\!\!\!\!\! {\rm d}t\,
    (t+1)^{\nu-\frac{1}{2}} (t-1)^{-\nu-\frac{1}{2}} \rme^{- z t/2}
 \bigg\}
  \bigg]
 \nnn
\fl
= 
 \frac{-1}{4\pi}\,
 \Gamma(\tfrac{1}{2}-\nu)\, z^{-\frac{1}{2}}\, W_{\nu,0}(z)
\end{eqnarray}
%
%
where we used a logarithmic representation of the inverse cotangens
and the reflection relation $\Gamma(\tfrac{1}{2}-\nu)
\Gamma(\tfrac{1}{2}+\nu) \cpn = \pi$.  The last equality in
\eref{eq:contint} holds since the expression in square brackets may be
deformed to the (complex conjugate of the) contour integral found in
\cite[eq. (5.1.6)]{Buchholz69}.  It gives the (real valued)
\emph{irregular Whittaker function} $W$ \cite{AS65} (multiplied by
$z^{-\frac{1}{2}}$) in an integral representation that is valid even for
positive $\nu$.
The regularized version of ${\rm G}^0$,
\begin{eqnarray}
  \Gtn_\nu(z) \defas
  \lim_{\mu\to\nu}\cos(\pi\mu)\,{\rm G}^0_{\mu}(z),
\end{eqnarray}
reads in terms of the more common \emph{irregular confluent hypergeometric
function} U 
\cite{AS65},
\begin{eqnarray}
  \Gtn_\nu(z) 
  = \frac{-1}{4\pi}\,\frac{\pi}{\Gamma(\nu+\tfrac{1}{2})}\,
  \rme^{\ts-z/2} \, {\rm U}(\ohmnu,1;z).
\end{eqnarray}
At small distances, it has the logarithmic form,
\begin{eqnarray}
  \Gtn_\nu(z) 
  =  {\rm L}_\nu(z) + {\rm O}(z\log z) \q\q\mbox{as $z\to 0$,}
\\
  \label{eq:defL}
  {\rm L}_\nu(z) \defas  \frac{\cpn}{4\pi}
      (\log(z)+{\Psi}(\ohpnu)-2{\Psi}(1) )
      - \frac{\sin(\pi\nu)}{4}
\end{eqnarray}
where $\Psi$ is the Digamma function.

\subsection*{A.1 The derivatives and their asymptotic behavior}

Employing the differential properties of the confluent hypergeometric
function \cite{AS65} we can express the derivatives of $\Gt^0_\nu$ by
the function $\Gt^0_\nu$ itself.
\begin{eqnarray}
\label{eq:zdG}
  z\frac{\rmd}{\rmd z}\Gtn_\nu(z) 
  &=& -(\ohmnu)(\Gtn_\nu+\Gtn_{\nu-1})-\frac{z}{2}\Gtn_\nu
  \\
\label{eq:zzddG}
  z^2\frac{\rmd^2}{\rmd z^2}\Gtn_\nu(z) 
  &=& (\tfrac{3}{2}-\nu)(\ohmnu)(\Gtn_\nu+2\Gtn_{\nu-1}+\Gtn_{\nu-2})
\nnn
  &&+z (\ohmnu) (\Gtn_\nu+\Gtn_{\nu-1})
  +\frac{z^2}{4}\Gtn_\nu
\end{eqnarray}
In section \ref{sec:solv} we need their asymptotic expansions,
\begin{eqnarray}
\fl
  z\frac{\rmd}{\rmd z}\Gtn_\nu(z) 
  &=& \frac{\cpn}{4\pi}
    \big[ 
       1 -z\,\nu\,
         \big(\log(z)+{\Psi}(\ohmnu)-2{\Psi}(1)-1\big)
    \big] 
  + {\rm O}(z^2\log z)
\\
\fl
  z^2\frac{\rmd^2}{\rmd z^2}\Gtn_\nu(z) 
  &=& - \frac{\cpn}{4\pi}   + {\rm O}(z\log z) \q\q\mbox{as $z\to 0$.}
\end{eqnarray}
These were deduced from the logarithmic representation of U in terms of
the regular Kummer function \cite[eq. (13.6.1)]{AS65}.

%
%
%
\setcounter{equation}{0}
\renewcommand{\theequation}{B.\arabic{equation}}
\section*{Appendix B. Analytical calculation of the singular integrals}
The Fourier integrals \eref{eq:defLl}, \eref{eq:defMl} depend on the
the window function $g$. Our choice is \eref{eq:defg} which switches
off the asymptotically singular functions m and l sufficiently
smoothly.  For the logarithmic integrals one obtains
\newcommand{\PP}{\Omega_l} \newcommand{\pp}{\varphi}
\begin{eqnarray}
\fl
   {\rm L}_\ell(s_0) = 
   \int_{-\srange}^{\srange}
   \!\!\!\!\!\!\rmd s' 
   \, \rme^{\rmi(2\pi \ell/\Len+\tfrac{\tvec_0\times\rvec_0}{b^2})s'}
   \Big(\rmi\Big[\alpha-\frac{s'}{b^2}\Big]\mp\lambda\Big[\frac{2\nu}{b^2}
     +(\alpha-\rmi\kappa_0)\frac{s'}{b^2}\Big]
   \Big)\,
  {\rm L}_\nu\Big(\frac{{s'}^2}{b^2}\Big)\,
  g(s')
  \nnn
  = 
  \big( \rmi\alpha\mp\lambda\frac{2\nu}{b^2}
  \big) {\rm I}_{\cos} 
  + \big( 1\mp\lambda(\kappa_0+\rmi\alpha)
    \big) {\rm I}_{\sin} 
\end{eqnarray}
with
\begin{eqnarray}
  \fl
   {\rm I}_{\cos} \defas
   \frac{\cpn}{4\pi} \frac{-1}{\PP\,\pp^+\,\pp^-}
  \Big\{
    \pi^2  \sin(\pp)
       \Big[ \log\Big(\frac{\srange^2}{b^2}\Big) 
             + {\Psi}(\ohmnu)-2{\Psi}(1) \Big]
\nnn
    + 2\pp^+\pp^- \, \Si(\pp)
    + \pp\,\pp^+  \, \Si(\pp^-)
    + \pp\,\pp^-  \, \Si(\pp^+)
    \Big\}
\\
  \fl
   {\rm I}_{\sin} \defas
   \frac{\cpn}{4\pi} \frac{1}{\PP^2 b^2\,(\pp^+)^2\,(\pp^-)^2}
  \Big\{
    \pi^2 \pp\, \pp^+ \pp^-  \cos(\pp)
       \Big[ \log\Big(\frac{\srange^2}{b^2}\Big) 
             + {\Psi}(\ohmnu)-2{\Psi}(1) \Big]
    \nnn
    -\pi^2(3\pp^2-\pi^2) \sin(\pp) 
            \Big[ \log\Big(\frac{\srange^2}{b^2}\Big)+ 2 
                  + {\Psi}(\ohmnu)-2{\Psi}(1) \Big]
\\
    -2 (\pp^+)^2(\pp^-)^2  \, \Si(\pp)
    - \pp^2(\pp^-)^2  \, \Si(\pp^+)
    - \pp^2(\pp^+)^2  \, \Si(\pp^-)
    \Big\}
    \nn
\end{eqnarray}
where $\PP(s_0)=2\pi \ell/\Len+\tfrac{\tvec_0\times\rvec_0}{b^2}$,
$\pp=\PP(s_0)\,\srange$, $\pp^\pm= \pp\pm\pi$,  and $\Si$ is the
Sine Integral.
The finite part integral reads
\begin{eqnarray}
\fl
   {\rm M}_\ell(s_0) = 
   \mp\lambda
   \frac{\cpn}{2\pi}
   \fpint_{-\srange}^{\srange}
   \!\!  \rmd s' 
   \,\, \rme^{\rmi(2\pi \ell/\Len+\tfrac{\tvec_0\times\rvec_0}{b^2})s'}
   \frac{-1}{{s'}^2}
   \cos^2\!\Big(\frac{\pi}{2}\frac{s'}{\srange}\Big)
\nnn
   = 
   \mp\lambda
   \frac{\cpn}{2\pi}
   \lim_{\epsilon\to0}
   \Big[
       2 \int_\epsilon^{\srange} 
         \cos(\PP s) 
         \frac{-1}{s^2}
         \cos^2\!\Big(\frac{\pi}{2}\frac{s}{\srange}\Big)
         \rmd s
       + \frac{2}{\epsilon}
   \Big]
\nnn
   = 
   \mp\lambda
   \frac{\cpn}{2\pi}
   \Big(
       \frac{1}{2\srange}
          \{  
              2 ( \cos(\pp) + \pp\, \Si( \pp) )
              + \cos(\pp^+) + \pp^+ \Si( \pp^+) 
\nnn
              + \cos(\pp^-) + \pp^- \Si( \pp^-) 
           \}
 + 
   \lim_{\epsilon\to0}
   \Big[
   - \frac{1}{2\epsilon} 
           \{  
              4 + \Or(\epsilon)
           \}
   +\frac{2}{\epsilon}
   \Big]
   \Big)
\nnn
   = 
   \mp\lambda
   \frac{\cpn}{4\pi\srange}
       \big\{  
             2  \pp\, \Si( \pp) + \pp^+ \Si( \pp^+) + \pp^- \Si( \pp^-) 
        \big\}.
\end{eqnarray}
Asymptotically,
\begin{eqnarray}
\label{eq:Masymp}
   {\rm M}_\ell(s_0) \sim \mp\lambda\, \frac{\cpn}{2}\, \PP(s_0)\, 
	{\rm sgn}(\ell)
   \qq
   \mbox{as $|\ell|\to\infty$.}
\end{eqnarray}
Note that with choice \eref{eq:defg} the limit of the remaining kernel
is
\begin{eqnarray}
\fl
  \lim_{s\to s_0}
  \Big[
    {\rm q}(s,s_0) - g(s-s_0)\,
             \big(     {\rm l}(s,s_0) + {\rm m}(s,s_0)
             \big) 
  \Big]
\nnn
  =
  \frac{\cpn}{4\pi}
    \Big[
      \kappa_0 (1\mp\lambda\rmi\alpha)
      \mp\lambda\big(-\frac{2\nu}{b^2}-\frac{\pi^2}{2\srange^2}\big)
    \Big]
\end{eqnarray}
which is \emph{not} just the constant part of \eref{eq:asympq} but contains a
term which depends on $\srange$.

%
%
%
\setcounter{equation}{0}
\renewcommand{\theequation}{C.\arabic{equation}}
\begin{figure}
\epsfxsize\textwidth%
\epsfbox{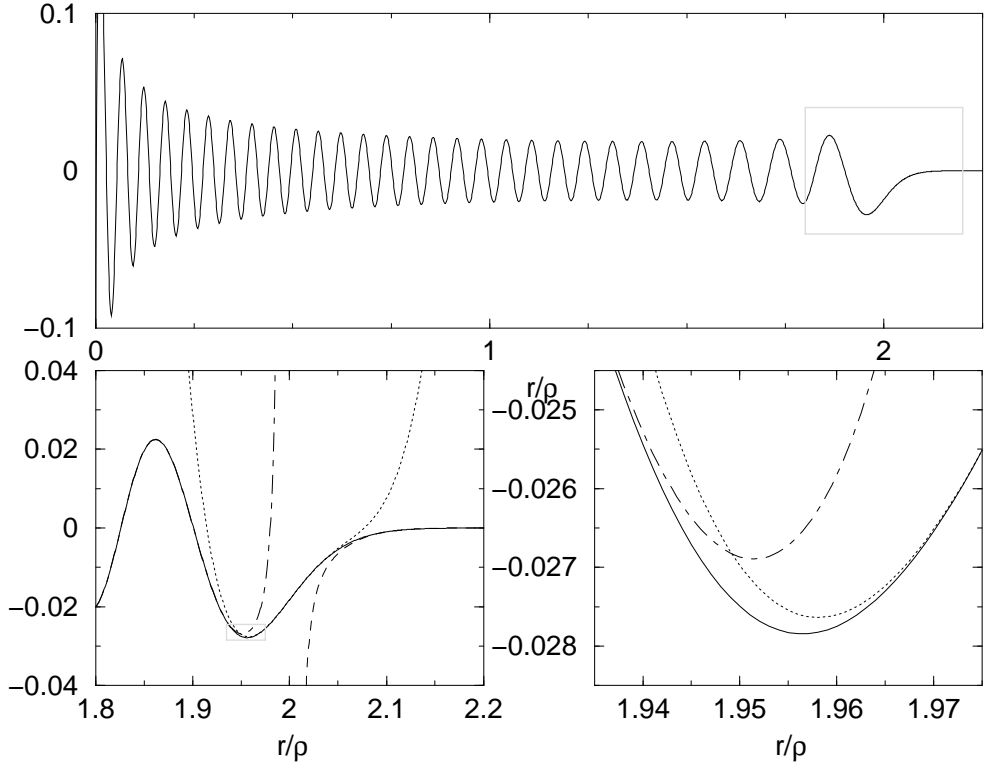}
\caption{%
(a) Gauge independent part of the regularized Green function at
$\nu=57.75$. It has a logarithmic singularity at $r=0$ 
and decays exponentially for $r>2\rho$.  (b) In the transition regions
between oscillatory, transient, and decaying regimes the asymptotic
expressions to third order are not valid (chain, dotted, dashed line
respectively.) (c) Here, on may interpolate using an uniform approximations
to the irregular Whittaker function (solid line.) }
\label{fig:greensfct}
\end{figure}

\section*{Appendix C. Numerical evaluation of the Green function}

We are not aware of any published numerical procedure to evaluate the
irregular confluent hypergeometric function ${\rm U}$ if both, the
(energy) parameter and the variable are large.  It seems that
presently only the Mathematica software (Wolfram Research Inc.) is
able to compute the function, at least for moderately large $\nu$.
Even this sophisticated system \emph{fails} for $\nu > 75$.  Anyhow,
it is not an option to use it for serious numerical calculations since
the evaluation takes a prohibitively long time.

Therefore, we describe our method to compute the gauge independent
part of the regular Green function in more detail.  For low energies
$\nu<12$, the function ${\rm U}(1/2-\nu,1;z)$ may be easily calculated
by its series representation \cite[eq. (13.1.6)]{AS65}. i.e. in terms
of the regular confluent hypergeometric function $_{1\!}{\rm
F}_{\!1}$. For very large $z$ an asymptotic expansion in terms of
$_{2}{\rm F}_{\!0}$ may be employed \cite[eq. (6.7.1)]{MOS66}.

For energies $\nu>12$ the numerical convergence of the series
expression deteriorates strongly in some intervals of the $z$ range
(starting at $z\approx 2\nu$).  Fortunately, a number of rather
complicated asymptotic expansions for the irregular Whittaker function
exist \cite[eqs. (8.1.5), (8.1.10), (8.1.18a)]{Buchholz69} which are
to \emph{third order} in the large parameter $\nu$.  These expressions
are based on saddle point approximations of a defining
integral. Together with \cite[eq. (13.5.15)]{AS65} they correspond to
the changing logarithmic, oscillatory, transient, and exponentially
decaying behavior of the Green function as the distance $z$ increases.
For most values of $z$ they allow to calculate the Green function to a
reasonably high precision and with acceptable numerical effort.
However, between the ranges of validity of the different asymptotic
expressions there are small gaps where no formula is appropriate,
cf. Figure \ref{fig:greensfct}.  In the gap between the logarithmic
and the oscillatory domains, which is at small $z$, one may employ the
series summation even for large $\nu\gg12$. For the two gaps between
the oscillatory, the transient, and the exponential regimes, which are
around $z\approx4\nu$ this is possible only up to, say $\nu=16$. For
larger $\nu$ we interpolate between adjacent regions of validity
employing the \emph{uniform approximation} of the irregular Whittaker
function around the classical turning point.  Neglecting higher orders
in $\nu$, the resulting expression for the Green function reads,
\begin{eqnarray}
  \Gt_\nu(z) \approx C\frac{(\frac{3}{2}q)^{\frac{1}{6}}}{
  |z^2-4\nu z-1|^\frac{1}{4}} 
 {\rm Ai}\Big(
	{\rm sgn}(z-z_0)\, \big(\tfrac{3}{2} q\big)^{\frac{2}{3}}
	\Big)
\end{eqnarray}
where Ai is the regular Airy function and
\begin{eqnarray}
\fl
  q=\Biggl\{
\begin{array}{ll}
\displaystyle
\nu\Big(
	\frac{\pi}{2}-{\rm atan}\Big(\frac{z-2\nu}{w}\Big)
	\Big)
    +\frac{1}{2}\log\Big(
		\frac{z_0}{z}\frac{1+2\nu z
                       +w}{1+2\nu z_0}
                    \Big)
    -\frac{1}{2}w
&\mbox{if $z<z_0$}
\\[-2ex]&
\\
\displaystyle
	\frac{1}{2}w + \frac{1}{2}{\rm atan}\Big(\frac{2\nu z +1}{w}\Big) 
	- \frac{\pi}{4}
	- \nu\log\Big(	\frac{z-2\nu+w}{z_0-2\nu} \Big)
&\mbox{if $z>z_0$}
\end{array}
\end{eqnarray}
\begin{eqnarray}
\fl
\mbox{with}\q
z_0=4\nu\, (\tfrac{1}{2}+\tfrac{1}{2}\sqrt{1+\tfrac{1}{4\nu^2}})
\q\mbox{and}\q
w=\sqrt{|z^2-4\nu z-1|}.
\end{eqnarray}
The constant $C$ may be calculated for values of $z$ where the saddle
point expressions are valid and is interpolated linearly within the gaps.

The thresholds mentioned above are a reasonable compromise between cost
and precision. We observe a peak numerical error (minimum of relative
and absolute) of $6.5\times 10^{-5}$ at $\nu=22$ by comparison with the
results of Mathematica which are assumed to be exact for
$\nu<70$. For increasing $\nu$ the numerical error decreases
monotonically which allows us to estimate it to smaller than
$3.7\times 10^{-5}$ for $\nu>70$. It was checked that numerical errors
of that order do not affect the results shown in section 6.

%
%
%

\end{document}